\documentclass[fleqn,10pt]{wlscirep}
\usepackage[utf8]{inputenc}
\usepackage[T1]{fontenc}
\usepackage{float}
\usepackage{multirow}
\usepackage{graphicx}
\usepackage{subfig}
\usepackage[justification=justified]{caption}
\usepackage{lipsum}
\usepackage[squaren,Gray]{SIunits}

\usepackage{hyperref}
\usepackage{soul}

\usepackage{amsmath}

\usepackage{tcolorbox}
\definecolor{darkred}{RGB}{150,0,0}
\definecolor{darkgreen}{RGB}{0,150,0}
\definecolor{darkblue}{RGB}{0,0,200}
\hypersetup{colorlinks=true, linkcolor=darkred, citecolor=darkgreen, urlcolor=darkblue}

\newcommand{\cln}[1]{\textcolor{red}{}}

\title{Dynamic Plastic Deformation Delocalization in FCC Solid Solution Metals}

\author[1,*]{Dhruv Anjaria}
\author[2]{Milan Heczko}
\author[3]{Daegun You}
\author[1]{Mathieu Calvat}
\author[1]{Shuchi Sanandiya}
\author[4]{Maik Rajkowski}
\author[4]{Aditya Srinivasan Tirunilai}
\author[3]{Huseyin Sehitoglu}
\author[4]{Guillaume Laplanche}
\author[1,*]{J.C. Stinville}

\affil[1] {Materials Science \& Engineering Department, University of Illinois Urbana-Champaign, Urbana, Illinois, USA} 

\affil[2] {Institute of Physics of Materials, Czech Academy of Sciences, Brno, Czech Republic} 

\affil[3] {Mechanical Science \& Engineering, University of Illinois Urbana-Champaign, Urbana, Illinois, USA} 
\affil[4] {Institut für Werkstoffe, Ruhr-Universität Bochum, Bochum, Germany}  

\affil[*]{corresponding author: Dhruv Anjaria (anjaria3@illinois.edu) and J.C. Stinville (jcstinv@illinois.edu)}

\begin{abstract}
Metallic materials undergo irreversible deformation under mechanical loading, leading to intense local plastic localization that reduces their mechanical performance. We identify a new mechanism of plastic deformation localization that dynamically promotes the homogenization of plasticity in face-centered cubic solid solution-strengthened metallic alloys. We observe that this mechanism occurs within a narrow range of stacking fault energies and involves competing deformation between nanoscale twinning and slip. This phenomenon is attributed to a new mechanism referred to as dynamic plastic deformation delocalization, which opens a new design space for enhancing the mechanical performance of metallic materials. We demonstrate that the activation of this mechanism has a significant impact on fatigue properties, greatly enhancing fatigue strength when it occurs.

\textbf{\\keywords: }{Plastic deformation delocalization, Fatigue, Solid solution metals, Mechanical performance, Nanotwinning}
\end{abstract}

\begin{document}

\flushbottom
\maketitle
\thispagestyle{empty}

% \noindent Please note: Abbreviations should be introduced at the first mention in the main text – no abbreviations lists. Suggested structure of main text (not enforced) is provided below.

\textit{This is a  \textbf{colorblind-friendly version} (protanopia, deuteranopia, tritanopia) following the guidance from Colour Blind Awareness \cite{Awareness}. The color-blind version of this article has been made possible thanks to funding from the NSF (award \#2338346).}

\hfill 

Metallic materials have critical importance since they are used in a vast range of structural applications\cite{senkov2004metallic} across a wide range of temperatures due to their excellent monotonic mechanical properties, including yield strength, ultimate tensile strength, ductility, and toughness \cite{SteelCryo,basuki2017alloys}. These properties originate from their atomic structure and engineered microstructure, which provide substantial strengthening and effectively accommodate defects and deformation \cite{doi:10.1126/science.1254581,Ritchie2011,OTTO20135743}. 

Their microstructure is often designed to exhibit significant heterogeneity, creating barriers to defect motion, such as dislocations, and promoting beneficial deformation mechanisms \cite{DING201921,Mishra2021}. Through this design approach, these microstructural barriers provide exceptional strength to the material \cite{DING201921}. However, this microstructure heterogeneity leads to highly localized plasticity at small scales, concentrating deformation within specific regions of the material \cite{stinville2022origins}. Consequently, these regions become highly susceptible to damage under cyclic loading (i.e. fatigue)\cite{MUGHRABI20131197,Mughrabi2009,Mughrabi2013, forrest2013fatigue}. 

As increasingly engineered microstructures are designed to exhibit unprecedented monotonic strength, the fatigue efficiency of these materials tends to decrease \cite{WANG2025120888,stinville2022origins,FLECK1994365}. Fatigue efficiency can be described by the fatigue ratio \cite{FLECK1994365}, defined as the fatigue strength normalized by the yield or ultimate tensile strength. When examining the fatigue ratio as a function of the yield or ultimate tensile strength for a large set of metallic materials, a significant decrease in the fatigue ratio is observed with increasing strength \cite{stinville2022origins,FLECK1994365}. In many cases, highly engineered materials with high strength can fail by fatigue at stresses as low as 25\% of their yield strength, highlighting a notably low fatigue efficiency. High-strength materials, such as superalloys \cite{10.1007/978-3-319-89480-5_2,gell1970fatigue} and titanium alloys\cite{Kang2019}, are examples of highly engineered materials that exhibit high strength but low fatigue efficiency. Therefore, fatigue remains a critical design consideration for structural materials.

This trade-off between monotonic strength and fatigue strength arises from the intense localization of plasticity and its irreversibility that occurs in metallic materials \cite{MUGHRABI20131197, stinville2020direct, stinville2022origins, POLAK2015386}. This relationship has been qualitatively demonstrated in metallic materials that deform by slip \cite{stinville2022origins}, where statistical measurements of slip localization intensity (i.e. plasticity localization induced by slip) have been related to the fatigue ratio. As slip localization intensity increases, the fatigue ratio (i.e., fatigue efficiency) decreases, and vice versa \cite{stinville2022origins}. In face-centered cubic (FCC) and hexagonal close-packed (HCP) materials, reducing the intensity of plasticity localization without compromising monotonic strength is challenging, as high yield strength FCC and HCP materials consistently exhibit intense plastic localization. This intense plastic localization is significantly pronounced in FCC materials at room temperature, where thermally activated cross-slip is unfavorable. In contrast, when cross-slip is promoted, such as at high temperatures for FCC materials or at room temperature for BCC materials, the intensity of plastic localization decreases significantly, and the fatigue ratio thus improves \cite{Ferro01101965, Buck1967}.
 
Massively promoting cross-slip in FCC materials to minimize the intensity of plastic localization is challenging and we are therefore seeking to identify either processing or microstructures that will provide low levels of plastic localization. One potential solution is to delocalize plastic deformation through conceptual mechanisms that lead to homogenized plasticity during deformation. It has been demonstrated as being possible in HCP materials by either identifying microstructures that homogenize plasticity or by activating mechanisms that promote plastic homogenization \cite{WU2020100675, Ahmadikia2024, ANJARIA2024120106}. However, no such mechanism or evidence of this effect on fatigue strength has been demonstrated for FCC materials, in which planar slip is often dominant and favors high plastic localization. 

Here, we detail a novel dynamic plastic deformation delocalization mechanism that significantly homogenizes plasticity during deformation in FCC solid-solution materials. This mechanism was observed to occur at low levels of plasticity and therefore relevant to fatigue properties. The underlying small-scale processes of this mechanism and its impact on fatigue strength are examined in detail. The influence of alloy chemistry is demonstrated, revealing a new design space that enables dynamic plastic homogenization, thereby improving fatigue efficiency without compromising monotonic strength. The proposed mechanism is analogous to the well-known dynamic Hall–Petch effect \cite{CHEN2024265,AN2021195,Kang2025}, which allows alloys to overcome the trade-off between monotonic strength and ductility in metals.

\subsection* {Plastic localization response in several FCC materials}

In order to characterize the plastic localization response of a material at the microstructural level and at low levels of deformation, a statistical and high-resolution approach was employed. High-resolution digital image correlation (HR-DIC) measurements were performed to analyze large fields of view on the order of a square millimeter with a nanometer resolution (i.e. 22 nanometers spatial resolution) to quantitatively capture the characteristics of each deformation event. This combination enables the precise statistical quantification of plastic deformation localization in metallic materials \cite{stinville2022origins,STINVILLE2020110600, BOURDIN2018307}. Details concerning HR-DIC measurements and mechanical testing procedures are provided in the Methods section and in the Supplementary Materials.

\begin{figure}
  \centering
 \includegraphics[width=1\textwidth]{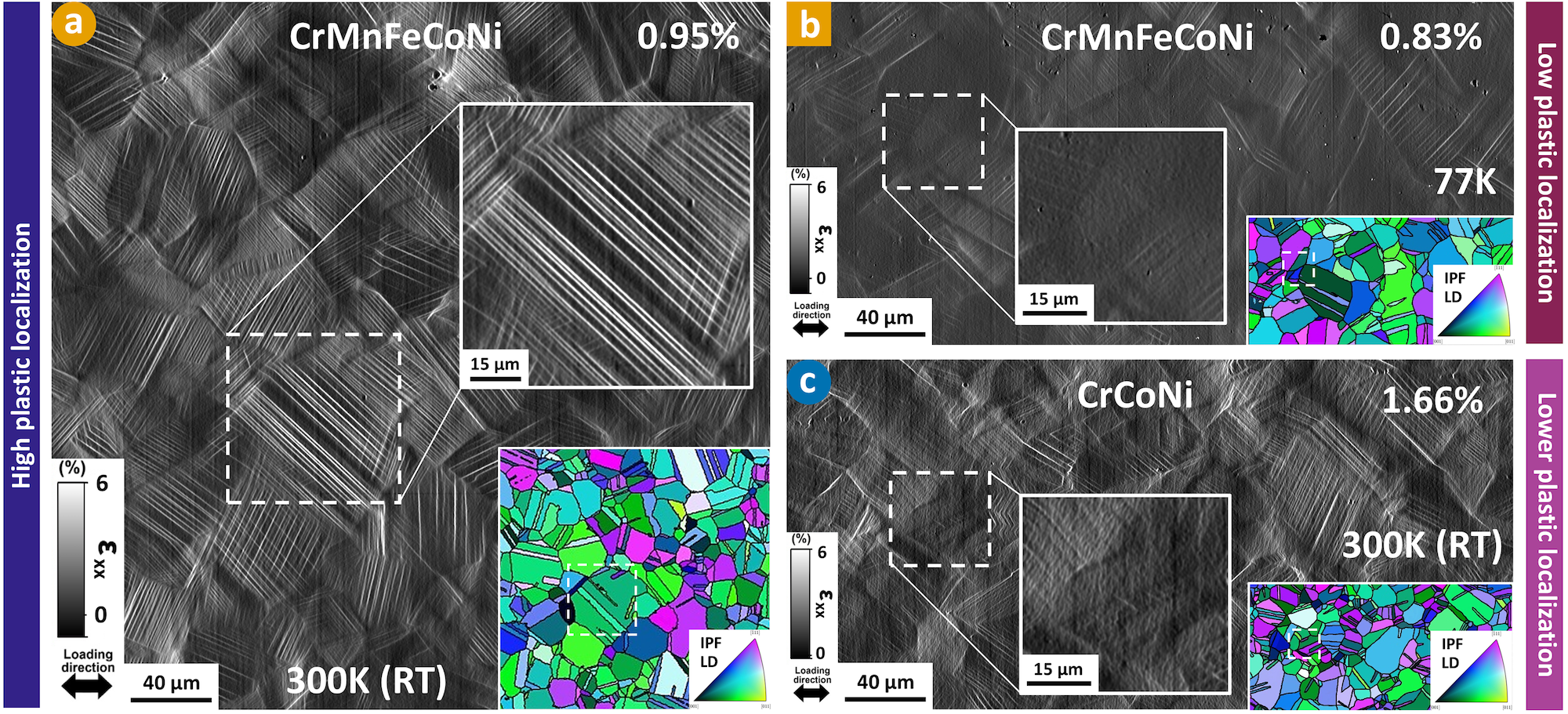}
\caption{\textbf{(a)} HR-DIC $\epsilon_{xx}$ longitudinal strain maps for CrMnFeCoNi deformed at room temperature (300K) up to a macroscopic plastic strain of 0.95\%, \textbf{(b)} CrMnFeCoNi deformed at liquid nitrogen temperature (77K) up to a macroscopic plastic strain of 0.83\% and \textbf{(c)} CrCoNi deformed at room temperature up to a macroscopic plastic strain of 1.66\%. The tensile direction is horizontal. The inverse pole figure maps along the loading direction obtained by electron backscatter diffraction are displayed in the insets. Reduced regions of interest are denoted using white dashed boxes and enlarged to highlight the differences in plastic deformation localization response across the investigated materials.}
\label{GrX_Maps}
\end{figure}

HR-DIC measurements were performed after unloading on several solid-solution-strengthened FCC materials, including CrCoNi, CrMnFeCoNi, FeNi36, VCoNi and stainless steel 316L, to quantify their plastic localization behavior at different loading temperatures, including room temperature (300K), liquid nitrogen (77K) and near-liquid helium temperature (20K). All materials were processed to be single-phase FCC solid solutions and to have similar microstructure in terms of grain size, grain shape, and texture. This ensures a direct comparison, as the intensity of plastic localization events is highly dependent on the microstructure \cite{YANG2025120682,JULLIEN2024146927,Mello2016, Alaie2015}. The details of these materials and their properties are provided in Table 1 in the Methods section.

%The mechanical properties of metallic materials are a function of the localization of plasticity that develops as a result of deformation. The intensity of plastic localization events directly affects properties such as fatigue strength, with highly intense events leading to a lower fatigue life\cite{}. 

%The intensity of these localization events depends on a number of factors. For instance, the grain size of the material affects localization of plasticity, and reducing the grain size leads to a lower intensity of plastic localization. A lesser known plastic response phenomenon is also emerging, which could be leveraged to improve the overall mechanical properties of a given material.

%The current work aims to reveal this phenomenon by analyzing several FCC solid solution strengthened alloys using HR-DIC after monotonic tensile testing at ambient and cryogenic temperatures. 

%The quantification of plastic deformation localization over respresentative microstructural regions is combined with site-specific conventional and high-resolution transmission electron microscopy to provide insights on the deformation mechanisms operating as a function of temperature and other material properties such as the stacking fault energy. 

We present the (HR-DIC) $\epsilon_{xx}$ longitudinal strain maps in Fig. \ref{GrX_Maps} for the CrMnFeCoNi and CrCoNi materials. The electron backscatter diffraction (EBSD) inverse pole figure (IPF) maps, showing the microstructure of these alloys corresponding to the representative regions, have been included in the insets. The IPF maps are colored along the loading direction (horizontal) using color-blind friendly coloring. The high entropy CrMnFeCoNi alloy, when deformed in tension at room temperature to a macroscopic plastic strain of 0.95\%, demonstrates intense plastic localization in the form of discrete deformation events as observed on the corresponding strain map in Fig. \ref{GrX_Maps}(a). This is further highlighted in the inset focusing on an individual grain within the investigated region. This plastic localization response, characterized by discrete intense deformation events (i.e. intense slip localization), is typically observed in FCC materials undergoing deformation at both room and cryogenic temperatures \cite{stinville2022origins, ANJARIA2024120106}. For most of the investigated materials and temperature conditions, we observed a similar response (see Supplementary Materials), consistent with previous reports in the literature \cite{stinville2022origins}. Surprisingly, for the CrMnFeCoNi deformed at liquid nitrogen temperature (77K) and CrCoNi deformed at room temperature to macroscopic plastic strains of 0.83\% and 1.66\%, respectively, we observed a drastically different plastic localization response. This plastic response was characterized by a homogeneous distribution of plasticity as evidenced by the strain maps presented in Fig. \ref{GrX_Maps}(b) and (c). Specific regions showing this homogenized plasticity response have been enlarged in the respective insets, indicating very low-intensity slip localization with plasticity extending over the entire crystallographic grains. The localization is remarkably uniform, rendering it impossible for the HR-DIC approach to capture individual deformation events within these materials. Nevertheless, high average strains are observed within each grain, indicating residual plastic deformation. Additional examples of this plastic response are shown in the Supplementary Materials.

Overall, an unusual plastic response in the form of homogeneous plasticity was observed in a few materials under specific testing conditions, thereby differing from the conventional localized plastic response typically exhibited by most investigated FCC metals.

\begin{figure}
  \centering
 \includegraphics[width=1\textwidth]{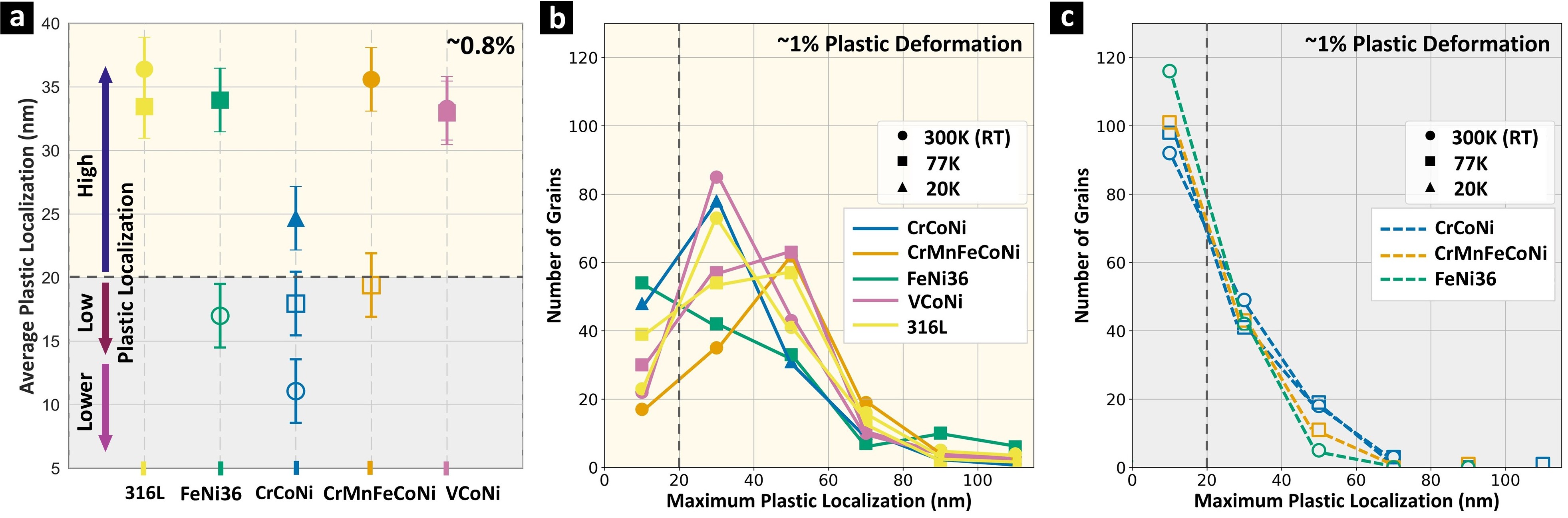}
\caption{\textbf{(a)} Average plastic localization in nanometers for several FCC solid-solution-strengthened alloys deformed up to a macroscopic plastic strain of 0.8\% at different temperatures. \textbf{(b,c)} Distribution of the number of grains as a function of the maximum plastic localization for the investigated materials. Materials that develop intense plastic localization are represented using solid lines and solid symbols \textbf{(b)}, whereas materials showing diffused, homogeneous plasticity are represented using dashed lines and open symbols \textbf{(c)}, showcasing a clear difference in the distribution trend between materials showing localized or diffused plasticity.}
\label{SFE_BlN_Grains}
\end{figure}

\subsection* {Quantitative measurement of plastic localization}

The average plastic localization intensity, measured in nanometers, is presented in Fig. \ref{SFE_BlN_Grains}(a) for the investigated FCC solid-solution-strengthened alloys. The materials were subjected to monotonic tensile tests up to a macroscopic plastic strain of approximately 0.8\%. The average plastic localization is a quantitative metric that describes the intensity of plastic localization from individual deformation events. It corresponds to the magnitude of the cumulative Burgers vector projected onto the specimen surface and is extracted using the HR-DIC approach \cite{BOURDIN2018307}, which is detailed in the Methods section. Each deformation event, such as slip or twinning, induces a local displacement (i.e., a step and in-plane displacement) at the specimen surface, which can be quantified using HR-DIC \cite{ANJARIA2024120106}. The intensity of the displacement along the event, representing the in-plane displacement induced by the deformation event, is measured in nanometers. All deformation events (thousands of events) were segmented, and the maximum amplitude along each event was extracted. The average plastic localization is then defined as the average value of the intensity of all detected events across the entire investigated region (about a square millimeter per material), that includes thousands of deformation events.

We observe that most of the investigated materials exhibit a high average plastic localization. In contrast, CrMnFeCoNi deformed at 77K, FeNi36 deformed at 300K and CrCoNi deformed either at 300K or 77K develop a surprisingly low average plastic localization, with values reduced by almost a factor of two compared to materials that exhibit high plastic localization. We arbitrarily divided the investigated materials into two categories, classified as "High" for the first one (average plastic localization above 20 nm) and "Low" and "Lower" for the second one, represented by yellowish and gray colors, respectively, in Fig. \ref{SFE_BlN_Grains}(a). This background color scheme and denomination will be maintained throughout the manuscript. For the medium entropy alloy CrCoNi, in particular, the average plastic localization at room temperature is significantly lower, approximately three times lower, compared to the other alloys that conventionally localize plasticity.

%This trend highlights the existence of a narrow range of stacking fault energies, represented using a gray band in Fig. \ref{SFE_BlN_Grains}(b) for which the investigated materials show an unusually low average plastic localization.

%hTe major differences in the HR-DIC strain maps for materials falling within and outside this stacking fault energy range further validate this observation, as previously detailed in Fig. \ref{GrX_Maps}. Materials outside this range show intense plastic localization (Fig. \ref{GrX_Maps}(a)), while materials within this range show homogeneous, diffused plasticity (Fig. \ref{GrX_Maps}(b) and (c)). Subsequently, a threshold of 20 nm has been defined in Fig. \ref{SFE_BlN_Grains}(a) to distinguish between high and low average plastic localization response.

\begin{figure}
  \centering
 \includegraphics[width=1\textwidth]{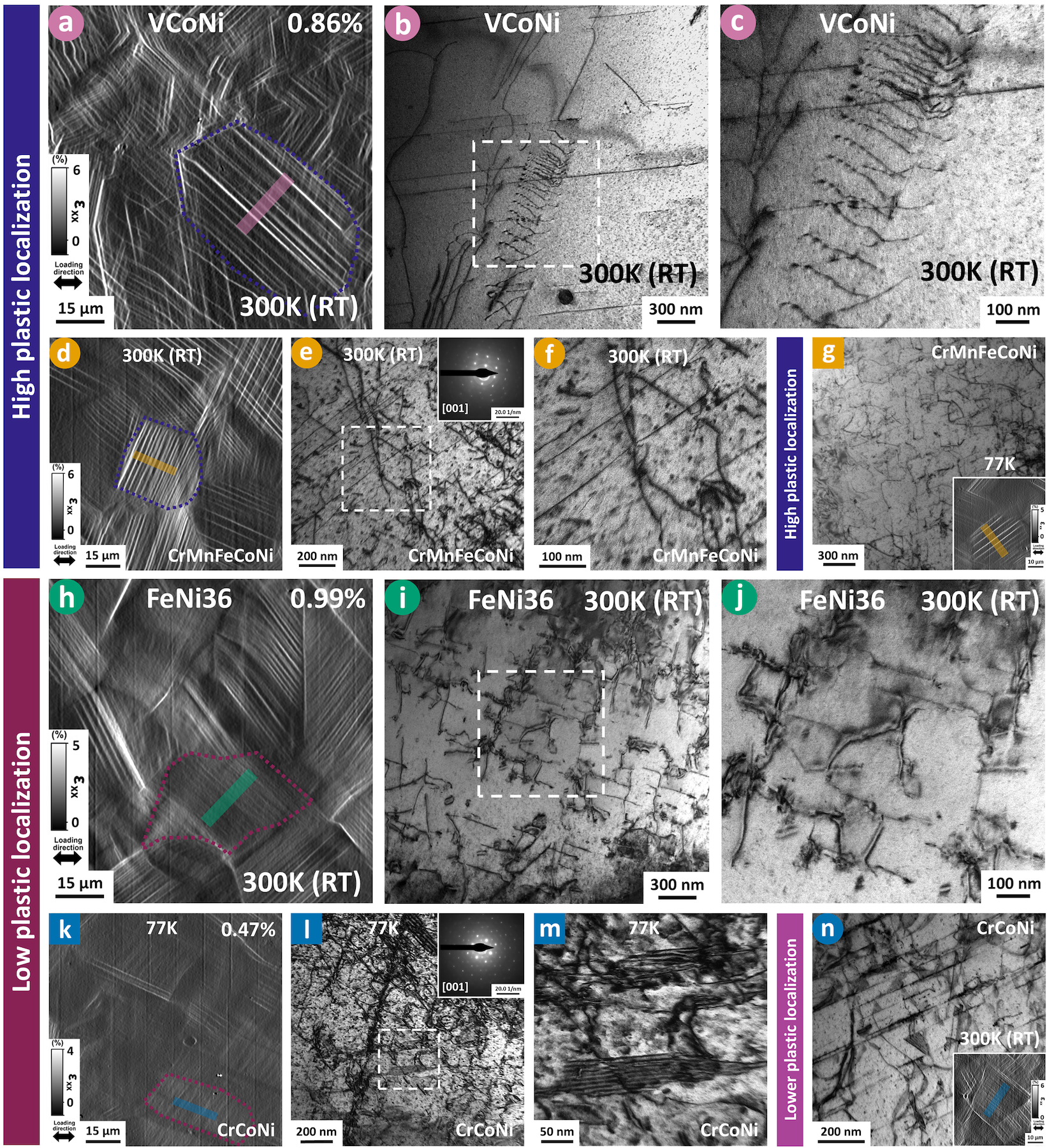}
\caption{Reduced regions of the HR-DIC $\epsilon_{xx}$ longitudinal strain maps for \textbf{(a)} VCoNi deformed at room temperature, \textbf{(d)} CrMnFeCoNi deformed at room temperature, \textbf{(h)} FeNi36 deformed at room temperature, and \textbf{(k)} CrCoNi deformed at 77K. The macroscopic plastic strains up to which the materials were deformed are included on the respective HR-DIC maps. Grains developing intense plastic localization and diffused plasticity were selected and TEM foils were extracted, marked as colored bands in \textbf{(a,d,h,k)}. Bright-field (BF) TEM images for the foils marked in \textbf{(a,d)} primarily reveal the presence of full dislocations across the foils \textbf{(b,e)}, whereas planar defects such as stacking faults and deformation twins are observed \textbf{(i,l)} for the foils marked in \textbf{(h,k)}, as evidenced by the SADPs in the insets. Regions marked with white dashed boxes in \textbf{(b,e,i,l)} have been enlarged in \textbf{(c,f,j,m)} respectively. Additional TEM foils extracted for different testing temperatures are shown in the insets in \textbf{(g)} and \textbf{(n)}. The corresponding BF TEM images show full dislocations and additional planar defects for localizing and diffused plasticity grains respectively \textbf{(g,n)}.}
\label{TEM_Combined}
\end{figure}

For each of the investigated materials at different testing temperatures, the distributions of the number of crystallographic grains as a function of the maximum plastic localization have been represented in Fig. \ref{SFE_BlN_Grains}(b) and (c). The materials were deformed up to a macroscopic plastic strain of around 1\%. The maximum plastic localization is expressed in nanometers and refers to the displacement induced by the most intense deformation event observed in a particular grain in the investigated representative microstructural region on the surface of the specimen. We observe from Fig. \ref{SFE_BlN_Grains}(b) that the materials that show intense plastic localization display a distinct trend compared to the materials that show homogeneous plasticity (i.e. low or lower plastic localization) represented in Fig. \ref{SFE_BlN_Grains}(c). For materials that develop intense localization, the distribution of the number of grains as a function of the maximum plastic localization shows peaks above our 20 nm threshold (see dashed black lines in Fig. \ref{SFE_BlN_Grains}). This displays that most of the considered grains in the investigated regions contain deformation events with a high plastic localization. On the other hand, materials exhibiting homogeneous plasticity (gray region in Fig. \ref{SFE_BlN_Grains}(a)) show a clear downward trend in the distribution of grains as a function of the maximum plastic localization. More importantly, they do not show grains with intense plastic localization. For these materials, most grains contain deformation events with extremely low plastic localization values (below our 20 nm threshold) or the plasticity is so diffused that individual deformation events cannot be resolved as discrete events using HR-DIC.

\subsection* {Transmission electron microscopy for characterization of defects}

We used HR-DIC $\epsilon_{xx}$ longitudinal strain maps to identify specific grains showing either intense plastic localization or homogeneous plasticity for the investigated materials and various testing temperatures. Thin foils were extracted from these locations and conventional transmission electron microscopy (TEM) analysis was performed as shown in Fig. \ref{TEM_Combined}. The strain maps corresponding to a reduced region of interest for the medium entropy alloy VCoNi and high entropy alloy CrMnFeCoNi deformed at 300K are shown in Fig. \ref{TEM_Combined}(a) and (d), respectively. Intense discrete deformation events are evident and the locations of the extracted TEM foils have been highlighted in the grains of interest using pink and orange bands. Bright-field (BF) TEM images primarily reveal the presence of full dislocations across extended areas of the foils, as observed in Fig. \ref{TEM_Combined}(b) and (e). Details of regions marked with a white dashed box are presented in Fig. \ref{TEM_Combined}(c) and (f), respectively. The selected area diffraction pattern (SADP) obtained by aligning the foil along the [001] zone axis reveals only diffraction spots corresponding to the FCC matrix as observed in the inset in Fig. \ref{TEM_Combined}(e). Although CrMnFeCoNi develops homogeneous plasticity when deformed at 77K (just below our threshold in Fig. \ref{SFE_BlN_Grains}(a)), the strain map shows grains with localized plasticity. Within this material, some grains demonstrate low plastic localization, while others show high plastic localization. Additional details on the distribution of the intensity of plastic localization are provided in the Supplementary Materials. A TEM foil was extracted from one grain with intense plastic localization and shown in the inset in Fig. \ref{TEM_Combined}(g). The corresponding BF TEM image in Fig. \ref{TEM_Combined}(g) reveals a distribution of full dislocations across the foil, similar to what was observed on the room temperature foil for this material. We followed the same approach to characterize the homogeneous plasticity ("Low" or "Lower" plastic localization) exhibited by some materials. The $\epsilon_{xx}$ longitudinal strain maps corresponding to FeNi36 deformed at 300K and CrCoNi deformed at 77K are shown in Fig. \ref{TEM_Combined}(h) and (k), respectively, along with the location of the extracted foils represented by green and blue bands. The observed plastic responses of the grains of interest differ drastically from the grains shown in Fig. \ref{TEM_Combined}(a) and (d) and are characterized by homogeneous plasticity. BF TEM images reveal the presence of dislocations across the foil along with a large number of planar defects such as stacking faults, as evidenced from Fig. \ref{TEM_Combined}(i) and (l). The regions marked with a white dashed box have been enlarged in Fig. \ref{TEM_Combined}(j) and (m), respectively to further highlight the presence of extensive stacking faults. In contrast to the SADP observed for CrMnFeCoNi at room temperature, the diffraction pattern observed in the inset in Fig. \ref{TEM_Combined}(l) reveals additional diffraction spots apart from the FCC matrix, indicating the presence of a high density of deformation twins, which we do not observe spatially on the BF TEM images. Please refer to the Supplementary Materials for a high-resolution view of the SADP. Therefore, it suggests the presence of a high density of nanometer-scale twins within this foil. Another TEM foil extracted from a grain with homogeneous plasticity (lower plastic localization) for CrCoNi deformed at 300K is shown in the inset of Fig. \ref{TEM_Combined}(n). The corresponding BF TEM image captured in Fig. \ref{TEM_Combined}(n) indicates the presence of dislocations as well as a high density of planar defects, similar to our observation on the foil corresponding to CrCoNi deformed at 77K, but with the addition of stacking fault tetrahedron \cite{LORETTO20151,KIRITANI1997133,Zhang2017}.

\begin{figure}
  \centering
 \includegraphics[width=1\textwidth]{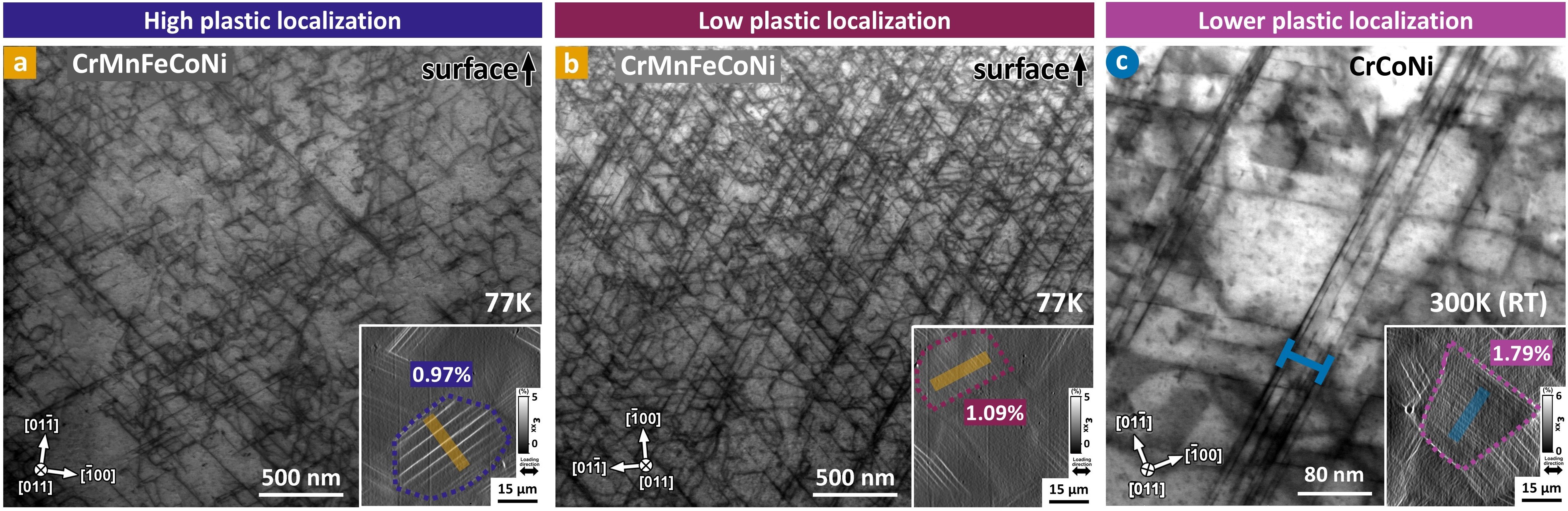}
\caption{BF-STEM DCI micrographs from \textbf{(a)} a (high) localizing grain for CrMnFeCoNi deformed at 77K, \textbf{(b)} a (low) delocalizing grain for CrMnFeCoNi deformed at 77K and \textbf{(c)} a (lower) delocalizing grain for CrCoNi deformed at 300K. The figure insets display reduced regions of the HR-DIC $\epsilon_{xx}$ longitudinal strain maps showing the grains from which thin foils were extracted along with the average strain in the considered grains. The electron beam is parallel with the [011] crystallographic zone axis of the investigated grains.}
\label{TEM2}
\end{figure}

Our conventional TEM analysis suggests that materials with homogeneous plasticity during deformation activate a high density of planar defects in addition to full dislocations. Conversely, materials with an intense plastic localization primarily exhibit full dislocations and conventional dislocation pile-up processes \cite{https://doi.org/10.1029/2005JB004021, CHAKRAVARTHY2010625}.

\subsection* {High-resolution transmission electron microscopy for defect type identification}

To further identify the types of defects in the investigated materials, multi-scale characterization using methods of scanning transmission electron microscopy (STEM) down to the atomic-resolution were performed on the different specimens. Detailed parameters used for STEM analysis are provided in the Methods section. A large field-of-view BF-STEM DCI micrograph from a grain exhibiting intense plastic localization in CrMnFeCoNi is shown in Fig. \ref{TEM2}(a). BF-STEM DCI micrographs from grains displaying homogeneous localization in CrMnFeCoNi and CrCoNi are shown in Fig. \ref{TEM2}(b) and (c), respectively. The strain maps associated with the grains from which the foils were extracted are included as insets, with the foil locations indicated by transparent orange or blue regions. The average strains along the loading direction ($\epsilon_{xx}$) of the investigated grains are also shown. In both the low and lower plastic localization cases, residual plastic deformation is present despite the absence of obvious intense deformation events.

Despite a similar average plastic strain in the two investigated grains in the high entropy alloy CrMnFeCoNi, the defect structures differ significantly. A higher density of defects, slip planes and dislocation debris are observed within the foil extracted from the grain with low plastic localization. In CrCoNi, characterized by the lowest plastic localization (see Fig. \ref{SFE_BlN_Grains}), the investigated grain shows homogeneous plasticity despite a high level of average plastic strain within the grain of 1.79\%. The deformation events display a high density of defects, slip planes and dislocation debris. Interestingly, they also exhibit an extended thickness, as shown in Fig. \ref{TEM2}(c) by the blue dimension line. Within this deformation event, we observed dislocation debris along multiple closely spaced slip planes. This specific structure was present in most of the deformation events within this foil.

\begin{figure}
  \centering
 \includegraphics[width=1\textwidth]{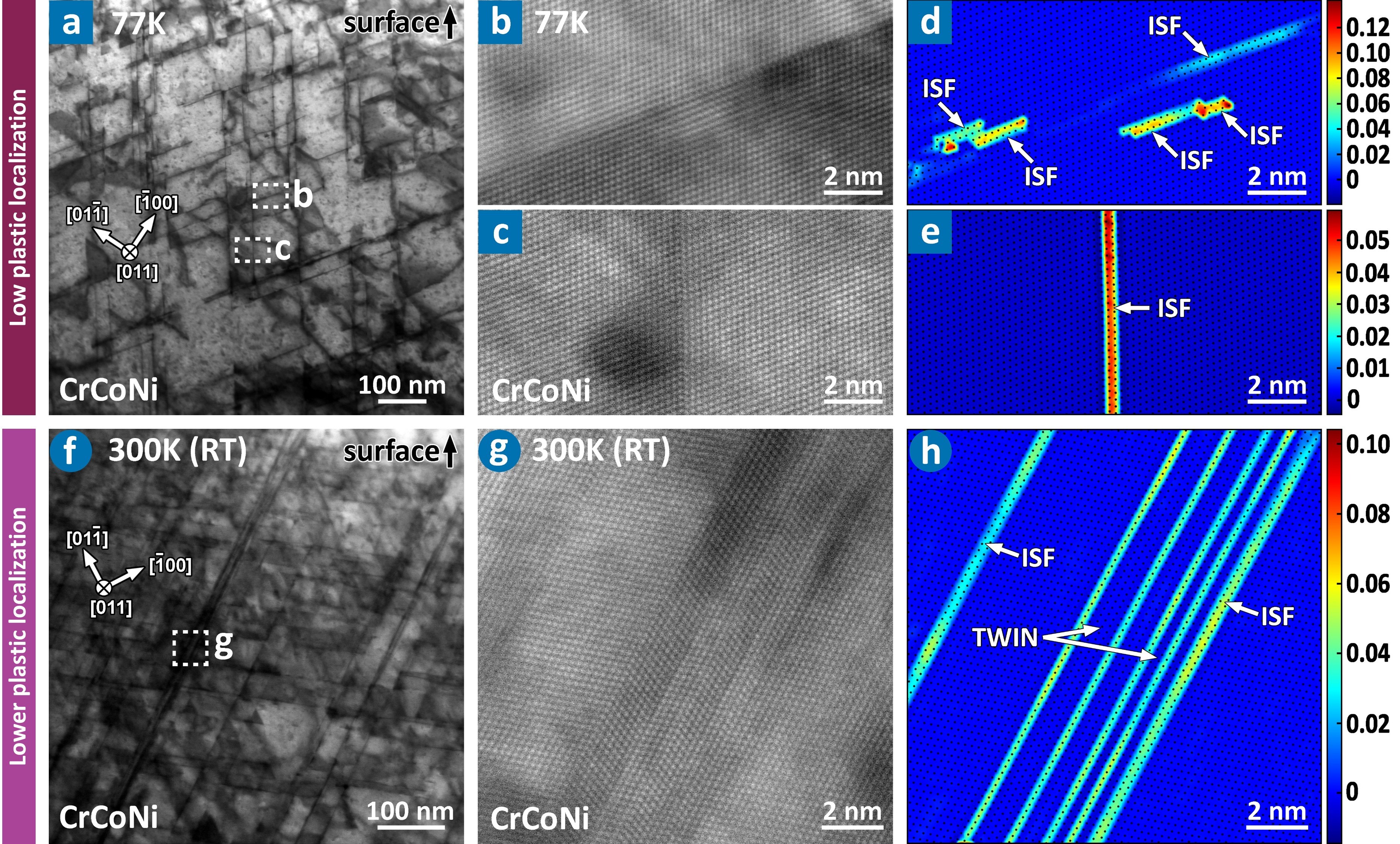}
\caption{\textbf{(a)} BF-STEM DCI micrograph from a (low) delocalizing  grain for the CrCoNi alloy deformed at 77K. \textbf{(b,c)} Atomic resolution HAADF-STEM images within two deformation events from (a). \textbf{(d,e)} Associated center of symmetry analysis. \textbf{(f)} BF-STEM DCI micrograph from a (lower) delocalizing grain for the CrCoNi alloy deformed at room temperature. \textbf{(g)} Atomic resolution HAADF-STEM image within a deformation event from (f). \textbf{(h)} Associated center of symmetry analysis. The electron beam is parallel with the [011] crystallographic zone axis of the investigated grains.}
\label{TEM3}
\end{figure}

Atomic-resolution HAADF-STEM images from within deformation events in the medium entropy alloy CrCoNi deformed at room temperature, shown in Fig. \ref{TEM3}(g), reveal the presence of several nanometer-scale deformation twins and other planar defects, such as intrinsic stacking faults. Center of Symmetry (COS) analysis \cite{HECZKO2021113985,PANDEY2023118928}, which measures the degree of centro-symmetry for each atomic column in the experimental HAADF-STEM image was performed to identify distortions in the stacking sequence. The results, displayed in Fig. \ref{TEM3}(h), confirm the presence of several nanometer-scale deformation twins within the observed deformation event in addition to intrinsic stacking faults. Additional examples provided in the Supplementary Materials demonstrate that this defect structure was consistently observed across all investigated deformation events in CrCoNi deformed at room temperature.

Similar imaging and analysis were conducted on CrCoNi deformed at 77K on a crystallographic grain which exhibits homogeneous plastic localization (low plastic localization), but with a slightly higher plasticity localization when compared to the same alloy tested at room temperature (lower plastic localization). In this case, as shown in Fig. \ref{TEM3}(d) and (e), within the deformation event, we detect only extended planar defects consisting of intrinsic and extrinsic stacking faults, hexagonal close-packed (HCP) structures, and stacking fault tetrahedra. However, no nanometer-scale deformation twins were detected. Finally, in the foil extracted from the grain exhibiting intense localization, no extended planar defects were identified within the deformation events.

\subsection* {A dynamic delocalization mechanism: the role of stacking fault energy}

\begin{figure}
  \centering
 \includegraphics[width=1\textwidth]{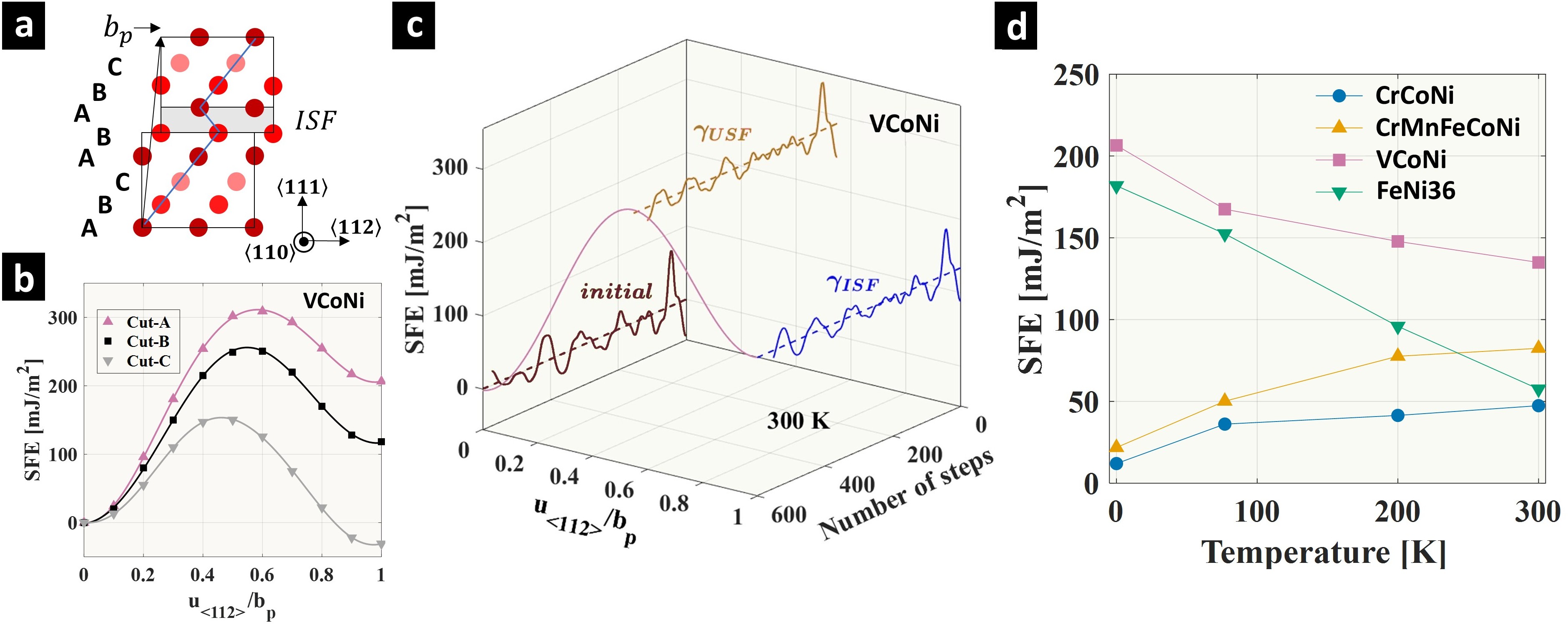}
\caption{\textbf{(a)} Schematic of stacking fault energy (SFE) calculation upon slip; Partial burgers vector (bp) along <112> direction is applied at intrinsic stacking fault (ISF) point; Stacking sequence (such as ABCABABC) is denoted after slip. \textbf{(b)} Example of SFEs in random VCoNi alloy at 0 K based on density functional theory (DFT); Different slip-cut configurations result in SFE variation. \textbf{(c)} Example of SFE of VCoNi alloy using ab-initio molecular dynamics (AIMD) at 300 K; The slip-cut where the maximum SFE is shown at 0 K is chosen (such as Cut-A in \textbf{(b)}). \textbf{(d)} SFE variation as a function of temperature for different alloys in this study. }
\label{SFECalculation}
\end{figure}

We calculated the stacking fault energies (SFE), which represent the energy per unit area associated with the formation of a stacking fault, as a function of temperature for CrCoNi, CrMnFeCoNi, FeNi36, and VCoNi. The SFE for stainless steel 316L was not calculated due to its complex atomic arrangement. The calculation method is detailed in the Methods section and is based on density functional theory (DFT) and ab-initio molecular dynamics (AIMD) to estimate stacking fault energies at 0K and finite temperatures, respectively. Fig. \ref{SFECalculation}(a) shows a schematic of an intrinsic stacking fault and the corresponding partial Burgers vector. An example of the SFE calculation at 0K is shown in Fig. \ref{SFECalculation}(b) for the medium entropy alloy VCoNi, highlighting the variation in calculated SFE values for different slip-cut configurations. To estimate the stacking fault energy at finite temperatures, the slip-cut corresponding to the maximum SFE at 0K is chosen, as shown in Fig. \ref{SFECalculation}(c). Finally, Fig. \ref{SFECalculation}(d) presents the calculated stacking fault energies as a function of temperature for the investigated materials. VCoNi\cite{YANG2021159} and FeNi36 both exhibit increasing stacking fault energy with decreasing temperature, whereas CrMnFeCoNi and CrCoNi show the opposite trend\cite{Fang2021}.

\begin{figure}
  \centering
 \includegraphics[width=1\textwidth]{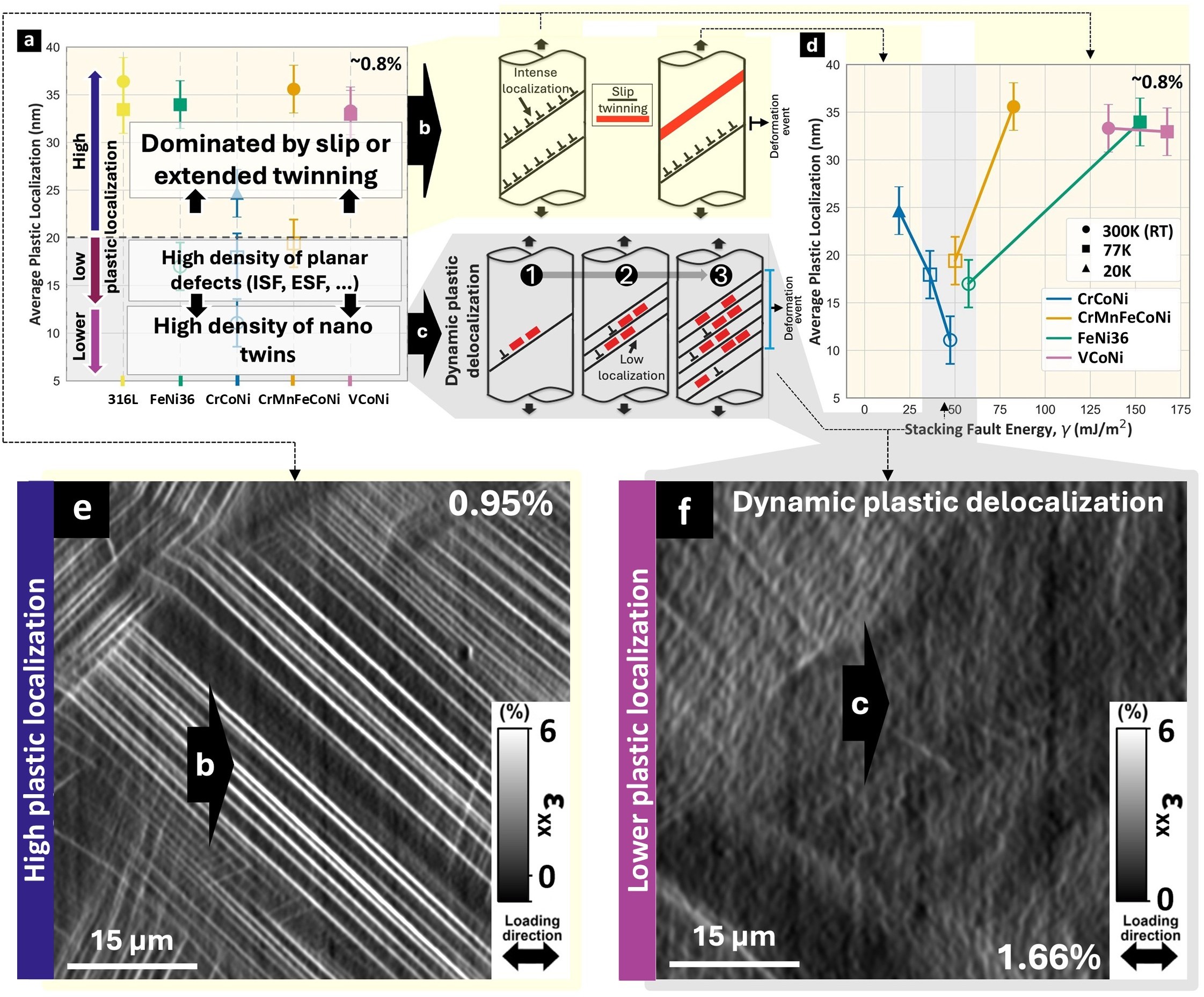}
\caption{\textbf{(a)} Observed variations in average plastic localization for several FCC solid-solution-strengthened alloys deformed at a macroscopic plastic strain of 0.8\%. Three distinct regimes were identified: one exhibiting intense plastic localization (labeled as "High"), another showing reduced plastic localization (labeled as "Low"), and a third characterized by a very low level of plastic localization (labeled as "Lower"). The latter two regimes are associated with the formation of planar defects, with the third regime further distinguished by a high density of nanometer-scale twins. \textbf{(b)} Conventional plastic localization processes involving either dislocation glide along specific slip planes or extended deformation twins were observed in the regime with intense plastic localization. \textbf{(c)} The proposed mechanism responsible for the reduced plastic localization amplitude involves the dynamic formation of nanometer-scale twins, which contribute to the thickening of deformation events. \textbf{(d)} The average plastic localization is shown as a function of the stacking fault energy of the alloys at the testing temperatures. The dynamic mechanism that results in a very low level of plastic localization occurs for alloys and temperature conditions within an intermediate range of stacking fault energies. \textbf{(e,f)} Examples of HR-DIC $\epsilon_{xx}$ longitudinal strain maps for intense plastic localization and homogeneous plasticity, respectively.}
\label{LastFigure}
\end{figure}

We summarize both the HR-DIC and S/TEM analyses in Fig. \ref{LastFigure}(a). We observe that the presence of nanometer-scale planar defects within deformation events significantly reduces plastic localization. In the case of nanometer-scale deformation twins, plastic localization is even more strongly suppressed. Above of our plastic localization threshold, we only observed the presence of full dislocations (high stacking fault energy) or long deformation twins (low staking fault energy case for the CrCoNi deformed at 20K and detailed in Supplementary Materials) that promote the usual intense plastic localization observed in metallic materials \cite{CHAKRAVARTHY2010625, stinville2022origins}. The simplistic schematic of the conventional deformation processes occurring in FCC materials at both room and low temperatures is illustrated in Fig. \ref{LastFigure}(b). These processes typically involve planar slip, for which a large number of dislocation events glide along a given crystallographic plane and emerge at the surface, thus leading to large surface steps and resulting in an intense localization \cite{CHAKRAVARTHY2010625, stinville2022origins}. Similarly, when deformation twinning dominates, large and extended twins form, producing significant surface steps associated with an intense localization. 

Fig. \ref{LastFigure}(d) shows the average plastic localization for the investigated materials and various temperature conditions as a function of the calculated SFE at each temperature. The background color scheme in Fig. \ref{LastFigure}(d) matches that of Fig. \ref{SFE_BlN_Grains}(a), where materials and testing conditions that exhibit low plastic localization (below our threshold) are shown on a gray background. We observe that all the materials and testing conditions exhibiting low plastic localization, indicative of plastic homogenization, fall within a narrow range of intermediate SFE values. For materials with high SFE, our S/TEM analyses revealed that deformation is primarily dominated by full dislocations. In contrast, for CrCoNi deformed at 20K (shown in the Supplementary Materials), we observed the formation of extended deformation twins. CrCoNi deformed at room temperature, which exhibits the lowest level of plastic localization, falls within the middle of the narrow region of intermediate SFE values. In this case, deformation events are characterized by large thicknesses (closely spaced slip bands) and contain a high density of nanometer-scale deformation twins and full dislocations. The underlying mechanisms are simplistically schematically illustrated in Fig. \ref{LastFigure}(c). 

These nanometer-scale twins, which were not present prior to deformation are observed to form during deformation and influence dislocation glide by forcing dislocations to glide on closely spaced crystallographic planes, as observed in Fig. \ref{TEM2}(c). This suggests a dynamic mechanism that promotes plastic homogenization through the competition and interaction between nanometer-scale deformation twins and full dislocation glide. We had initially observed evidences of this plastic homogenization phenomenon through closely-spaced slip in a previous study investigating the precipitation strengthened nickel based superalloy Inconel 718 at ambient and cryogenic temperatures \cite{ANJARIA2024120106}. At cryogenic temperatures (77K and 9K), deformation events consisted of multiple closely spaced slip traces, with detailed atomic resolution HAADF-STEM analysis revealing the presence of nanoscale planar defects such as intrinsic and extrinsic stacking faults, twins and HCP regions in the vicinity of these multiple slip events. 

The observed behavior may be explained by two potential mechanisms. The first involves the role of planar defects and their associated atomic arrangements in promoting double cross-slip, as suggested in reference\cite{NOHRING2017135}. Interestingly, the plastic delocalization revealed in our study occurs in materials with relatively low stacking fault energy, challenging the conventional understanding that lower stacking fault energy suppresses cross-slip and favors planar slip. This suggests that the presence of these planar defects may be essential in enabling double cross-slip. A second possible mechanism involves the activation or deactivation of dislocation sources. A recent study on CrCoNi under room-temperature compression \cite{Xie2025} demonstrated that the formation of extended stacking faults in FCC materials can deactivate dislocation sources. Atomistic simulations showed that when the leading partial dislocation has a higher Schmid factor than the trailing one, it glides first, forming a stacking fault. However, the trailing partial remains inactive, resulting in an unstable configuration that increases the energy barrier for further glide. This leads to source deactivation after emitting a single partial, deviating from the conventional Frank-Read dislocation multiplication model \cite{PhysRev.79.722}. Moreover, twin boundary migration induced by partial dislocation glide may create additional dislocation sources at stacking fault–twin boundary interfaces. This will promote dislocation glide on successive planes, ultimately forming the closely spaced slip bands we observed.

\subsection* {A new design space for enhancing fatigue strength}

\begin{figure}
  \centering
 \includegraphics[width=1\textwidth]{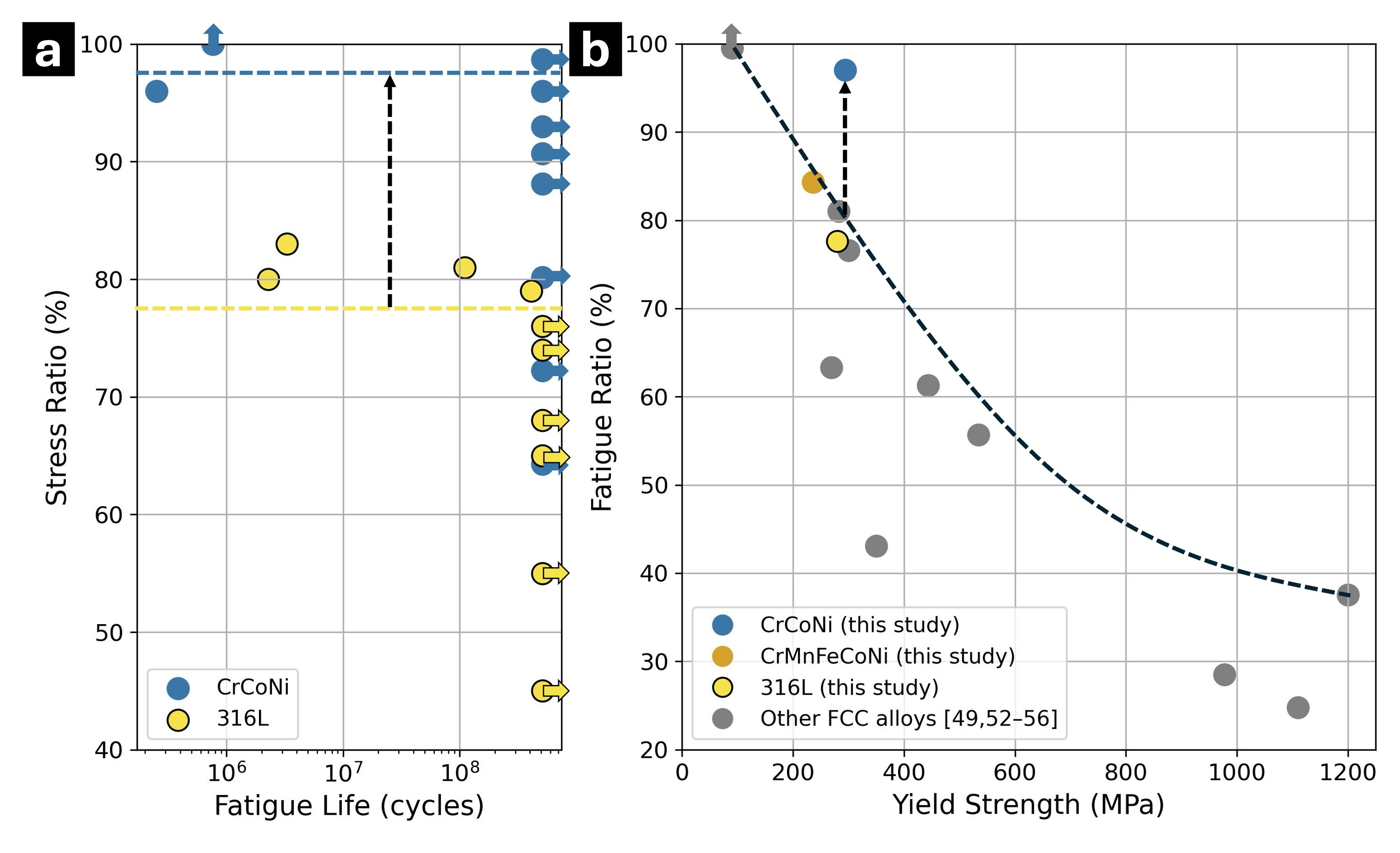}
\caption{\textbf{(a)} Fatigue life at room temperature of the investigated CrCoNi, CrMnFeCoNi alloys, and 316L stainless steel as a function of the maximum applied stress in the very high cycle fatigue regime. The maximum stress is expressed as a percentage of the yield strength. \textbf{(b)} Fatigue strength at room temperature, reported as a percentage of the yield strength (fatigue ratio), plotted against the yield strength for the investigated alloys and additional FCC alloys reported in the literature~\cite{CHEN20051227,KOLYSHKIN2016272,STOCKER20115288,GHOMSHEH2020139034,STOCKER20112,MARTI2020138619}.
Horizontal arrows indicate specimens that did not fail. Vertical arrows indicate cases where the fatigue life or fatigue ratio exceeds the yield strength of the material.}
\label{Fatigue}
\end{figure}

The influence of the identified dynamic plastic deformation delocalization mechanism on fatigue properties was evaluated by measuring the fatigue strength of CrCoNi, CrMnFeCoNi, and 316L stainless steel at room temperature in the very high cycle fatigue (VHCF) regime. Stainless steel served as a reference material due to its conventional plastic localization behavior. Among the tested alloys, CrMnFeCoNi exhibits the highest degree of plastic localization, while CrCoNi shows the lowest. The tendency of an FCC material to localize plastic deformation is known to correlate with its fatigue performance~\cite{stinville2022origins}; materials exhibiting intense localization typically demonstrate lower fatigue ratios, defined as the fatigue strength divided by the yield strength, following the work proposed by Fleck, Kang, and Ashby~\cite{FLECK1994365}. In agreement with the consideration that low plastic localization enhances fatigue properties, prior studies have reported superior fatigue performance for CrCoNi compared to CrMnFeCoNi at room temperature~\cite{LU2021113667,GHOMSHEH2020139034}. However, these comparisons must be contextualized, as those studies primarily addressed low-cycle fatigue (LCF) under relatively high strain amplitudes, where alternative deformation structures such as dislocation cells and vein patterns may dominate the fatigue response~\cite{LI2011328, PHAM2013143}. In the present study, fatigue performance in the VHCF regime is further compared with literature data for recrystallized FCC alloys tested under similar conditions~\cite{CHEN20051227,KOLYSHKIN2016272,STOCKER20115288,GHOMSHEH2020139034,STOCKER20112,MARTI2020138619}. The fatigue life of CrCoNi, CrMnFeCoNi, and 316L stainless steel at room temperature, plotted as a function of maximum applied stress (normalized by their yield strength), is shown in Fig. \ref{Fatigue}(a). Horizontal arrows indicate specimens that did not fail. CrCoNi demonstrates significantly higher fatigue strength compared to CrMnFeCoNi and 316L stainless steel. Figure \ref{Fatigue}(b) displays the measured fatigue ratios as a function of the yield strength. A general trend of decreasing fatigue ratio with increasing yield strength is typically observed for FCC metals at room temperature~\cite{stinville2022origins}. However, CrCoNi appears as a clear positive outlier, exhibiting a significantly higher fatigue ratio than expected. These findings highlight the beneficial role of dynamic plastic deformation delocalization in enhancing fatigue properties and suggest a transformative design strategy for the development of fatigue-resistant alloys.

\hypertarget{Methods}{}
\section*{Methods}\label{sec:Methods}

%\subsection*{Quantitative and statistical measurements of plastic localization}

%Place holder for characterization details.

\subsection*{Materials}

Five different face centered cubic (FCC) solid-solution-strengthened alloys were investigated in this study, namely CrCoNi, CrMnFeCoNi, VCoNi, FeNi36 and stainless steel 316L. For CrCoNi, CrMnFeCoNi and VCoNi, the pure elements were vacuum induction melted and cast into ingots. The ingots were homogenized at 1200$^{\circ}$C for 48 hours, followed by rotary swaging for diameter reduction. Recrystallization heat treatments were carried out at 1080$^{\circ}$C for 1 hour, 1060$^{\circ}$C for 1 hour and 1100$^{\circ}$C for 15 minutes for CrCoNi, CrMnFeCoNi and VCoNi respectively, after which tensile specimens were extracted. 316L specimens were extracted from a water quenched rolled plate, subjected to an annealing treatment at 1055$^{\circ}$C for 1 hour under high vacuum, followed by water quenching. FeNi36 specimens were taken from a forged plate subjected to an annealing treatment at 825$^{\circ}$C for 90 minutes followed by air cooling. The chemical composition of the investigated alloys in atomic percent along with the grain size after processing are listed in Table 1. The grain size includes annealing twins.

\begin{table}[h!]
    \centering
    \setlength{\tabcolsep}{4pt}
    \caption{Chemical composition (at.\%), grain size (GS) and yield strengths (YS) of the alloys considered in this study.}
    \label{tab:materials}
    \begin{tabular}{|c||c|c|c|c|c|c|c|c|c|c|c||c||c|}
        \hline
        \textbf{Alloy} & Co & Cr & Ni & Fe & Mn & V & Al & Si & Ti & Mo & C  & GS & YS (RT/LN$_2$/LHe)\\
        \textbf{Denomination} & ($\%$) & ($\%$) & ($\%$) & ($\%$)  & ($\%$) & ($\%$) & ($\%$) & ($\%$) & ($\%$) & ($\%$) & ($\%$)  & ($\mu$m) & (MPa)\\
        \hline
       CrCoNi & 33.33 & 33.33 & 33.33 & / & / & / & / & / & / & / & /  & 25 & 300/500/630 \\
        \hline
        CrMnFeCoNi & 20 & 20 & 20 & 20 & 20 & / & / & / & / & / & /  & 32 & 200/395 \\
        \hline
        VCoNi & 33.33 & / & 33.33 & / & / & 33.33 & / & / & / & / & / & 30 & 600/780 \\
        \hline
        FeNi36 & 0.48 & 0.27 & 36.59 & 60.32 & 0.61 & / & 0.21 & 0.81 & 0.12 & / & 0.23 & 27 & 290/630 \\
        \hline
        316L & 0.18 & 18.71 & 11.24 & 65.67 & 1.71 & / & / & 0.62 & / & 1.31 & 0.1  & 50 & 280/630\\
        \hline
        %NiFeCrNb & / & 21.21 & 54.56 & 18.97 & / & / & 1.27 & / & 1.28 & / & / & 2.42 & 25 & 370/540\\
        %\hline
    \end{tabular}
\end{table}

%\begin{table}[h!]
%    \centering
%    \setlength{\tabcolsep}{4pt}
%    \caption{Macroscopic plastic strains and yield strengths of the alloys considered in this study.}
%    \label{tab:materials}
%    \resizebox{1.05\textwidth}{!}{%
%    \begin{tabular}{|c||c|c|c|c|c|c|c|c|c|c|c|c|c|c|}
%        \hline
%        \ & CrCoNi & CrCoNi & CrCoNi & CrMnFeCoNi & CrMnFeCoNi & VCoNi & VCoNi & FeNi36 & FeNi36 & 316L & 316L  & NiFeCrNb & NiFeCrNb  \\
%        \ & (RT) & (LN2) & (LHe) & (RT)  & (LN2) & (RT) & (LN2) & (RT) & (LN2) & (RT) & (LN2) & (RT) & (LN2) \\
%        \hline
%       Macroscopic & 1.26 & 0.46 & 0.84 & 0.28 & 0.19 & 0.55 & 0.59 & 0.24 & 0.19 & 0.26 & 0.34 & 0.69 & 0.54\\
 %       plastic strain & 1.66 & 0.73 & & 0.95 & 0.83 & 0.86 & 0.91 & 0.59 & 0.52 & 0.79 & 0.86 & & \\
 %       (\%) &  & 0.98 &  & &  &  & & 0.99 & 1.08 & & & & \\
 %       \hline
%        YS (MPa) & 300 & 500 & 630 & 200 & 395 & 600 & 780 & 290 & 630 & 280 & 630 & 370 & 540 \\
%        \hline
%    \end{tabular}%
%    }
%\end{table}

\subsection*{Electron microscopy}

Scanning electron microscopy (SEM) characterization was performed using a Thermo Fisher Scientific\texttrademark~ Scios 2 Dual Beam SEM/FIB with a field emission gun (FEG). The imaging was conducted over areas with dimensions of \unit{723x405}{\micro\meter}$^2$, using a 6$\times$5 grid pattern with 15~\% overlap between neighboring images. The imaging was performed at an accelerating voltage of \unit{10}{kV} and a current of \unit{0.80}{\nano\ampere}. Secondary electron images were acquired with a dwell time of \unit{10}{\micro\second} per pixel, a horizontal field of view of \unit{138}{\micro\meter} and at a working distance of \unit{5}{\milli\meter}. The resulting images have a resolution of 6144~px $\times$ 4096~px. Automated imaging was performed using the AutoScript software from ThermoFisher\texttrademark~ and a custom Python routine for automated focus/astigmatism and grid imaging before and after deformation. The Python routine has been made available on GitHub (\url{https://github.com/cmbean2/Automated-SEM-Procedure}). 

Electron backscatter diffraction (EBSD) scans were conducted at a 70$^{\circ}$ tilt using a step size of 1 micron, square grid collection, 200 frames per second acquisition, an accelerating voltage of \unit{30}{kV} and a current of \unit{6.4}{\nano\ampere}. All EBSD scans were performed prior to deformation on the same regions investigated by HR-DIC. EBSD was performed on a ThermoFisher Scientific Scios 2 Dual Beam SEM/FIB using a Hikari Super EBSD detector. EBSD maps were then adjusted by distortion to match HR-DIC with pixel resolution using the method described in reference \cite{CHARPAGNE2020110245}.

The deformation processes and the profiles of slip step that develop during deformation on the surface were observed and documented with the underlying microstructure using site- and orientation-specific TEM surface lamella preparation by focused ion beam (FIB). A ThermoFisher Scientific Helios SEM-FIB Dual-Beam system was used to extract the thin foils with a final cleaning at 5 kV and a current of \unit{16}{\pico\ampere}. A ThermoFisher Scientific Talos F200X G2 Scanning Transmission Electron Microscope (S/TEM) was used to capture conventional TEM images of the different defects developing in these foils. Microstructure and defect analysis from these foils down to atomic resolution were performed using image aberration-corrected and monochromated ThermoFisher Scientific Titan-Themis STEM at an acceleration voltage of 300 kV. STEM diffraction contrast imaging (DCI) was performed with bright (BF) and high-angle annular dark-field (HAADF) detectors by selection of the appropriate camera length \cite{Phillips2011}. Atomic resolution of the microstructure was performed by tilting the thin foils into specific low-index crystallographic zones. Data was collected and processed using ThermoFisher Scientific Velox software. STEM micrographs were corrected for potential sample drift and scanning beam distortions using the drift-corrected frame integration function of Velox. Center of Symmetry (COS) analysis, which determines the degree of centro-symmetry for each atomic column in the experimental HAADF-STEM image and thus identifies distortions in the stacking sequence, was performed according to the procedure described in \cite{HECZKO2021113985,PANDEY2023118928}. The differentiation between extrinsic stacking faults (ESF) and twin stacking is based on the description used in reference \cite{Niu2018}.

\subsection*{High-Resolution Digital Image Correlation}

Before deformation, a speckle pattern consisting of \unit{60}{\nano\meter} silver nanoparticles was formed on the surface of the sample, following the procedure developed by Montgomery et al.\cite{Montgomery2019}. HR-DIC was performed upon completion of each step using proprietary software, XCorrel \cite{BOURDIN2018307}. HR-DIC images were captured before loading and after unloading for subsequent analysis. The SEM images (6144~px $\times$ 4096~px) were divided into subsets of $31 \times 31$ pixels ($\unit{700}{\nano\meter} \times \unit{700}{\nano\meter}$) with an overlap of 28 pixels between each subset (i.e., step size of 3 pixels ($\unit{67}{\nano\meter}$)). A representative scanning electron image of the speckle pattern has been included in the Supplementary Materials. Heaviside digital image correlation (H-DIC) is used when considering displacement fields that involve discontinuous displacement, like those produced by slip localizations \cite{Valle2017Crack}. Compared to conventional DIC, H-DIC provides higher displacement resolutions below 10 nm and the ability to find multiple discontinuities within a subset. Here, H-DIC enabled direct measurement of the amplitudes of the slip localizations \cite{STINVILLE2020110600, BOURDIN2018307}. 

\subsection*{Mechanical testing}

Flat dogbone-shaped specimens with a gauge section of $1 \times 3$ ${mm}^{2}$ were machined by electrical discharge machining (EDM) with gauge lengths of either 6 ${mm}$ or 10 ${mm}$. These samples were mechanically tested on a NewTec Scientific$\texttrademark$ MT1000 5000 Newton tensile stage and an electromechanical 3R$\texttrademark$ VHCF MEG20TT frame equipped with a custom setup for cryogenic testing. The cooling system consists of a vacuum double-walled chamber, multilayer insulation material, and helium/nitrogen liquid container. The system is designed as a closed-loop to recycle helium. For tests conducted at cryogenic temperatures, specimens were exposed to either liquid helium or nitrogen pushed from a dewar using helium or nitrogen gas, respectively. The temperature measurement was conducted using a silicon diode located near the specimen that displays an accuracy of less than 1 Kelvin within a range of 1.4 to 500 Kelvin. The monotonic tensile tests were conducted at a quasi-static strain rate between $10^{-5}$~s$^{-1}$ to $10^{-4}$~s$^{-1}$ and were interrupted at multiple plastic strain values. The 0.2\% offset yield strength values for all alloys as a function of temperature are reported in Table 1 and the specimen geometry along with the testing setup have been shown in the Supplementary Materials.

\subsection*{Fatigue testing}

Very high cycle fatigue (VHCF) testing was performed on the medium entropy CrCoNi alloy, the high entropy CrMnFeCoNi alloy and the 316L stainless steel. For the CrCoNi and CrMnFeCoNi, the pure metals were vacuum induction melted and cast, after which the as-cast ingots were homogenized at 1200$^{\circ}$C for 48 hours. Rotary swaging was used to reduce the diameter of the homogenized ingots from 40 mm to 10.7 mm. Cylindrical fatigue specimens with a gauge diameter of 3 mm and length of approximately 70 mm were machined from these rods and subjected to an annealing treatment at 1080$^{\circ}$C for 1 hour for CrCoNi and at 1060$^{\circ}$C for 1 hour for CrMnFeCoNi, followed by air cooling to obtain an average grain size of around 30 $\mu$m. The specimens were subjected to mechanical polishing and finished with a 3 $\mu$m diamond suspension prior to testing. A 3R$\texttrademark$ VHCF MEG20TT ultrasonic fatigue testing frame was used to conduct the fatigue tests at room temperature. The specimens were designed to have their first tension-compression eigenmode at 20 kHz\cite{StanzlTschegg2014}. Potential self-heating  was eliminated by employing the "pulse-pause" method\cite{Stanzl1981}, corresponding to 300 ms of cycling followed by 700 ms of rest. The tests were performed under fully reversed tension-compression loading, corresponding to a fatigue ratio of $R = -1$. The fatigue strength was determined at $5\times10^{8}$ cycles by adopting the staircase approach\cite{Stcker2017}. Low levels of alternating stress were initially applied after which power increments of 3\% were made, corresponding to a stress increase of approximately 25 MPa. For any given stress level, if specimen failure did not occur after $5\times10^{8}$ cycles, the following test was conducted at an elevated stress level until failure. The general specimen geometry and the testing setup are shown in the Supplementary Materials.

\subsection*{Stacking Fault Energy calculation}

Theoretical approach to stacking fault energy (SFE) calculation involves density functional theory (DFT) at 0 K and ab-initio molecular dynamics (AIMD) at finite temperatures via Vienna Ab-initio Simulation Package (VASP) \cite{PhysRevB.54.11169}. First, the SFE calculation in DFT is performed following the atomic relaxation and determination of lattice constants for superlattice. We construct 270 atomic sites in the oriented lattice, aligning 9 $\{111\}$ planes into z-direction. These 270 atomic sites are substituted by different constituent elements based on each alloy composition. The atomic configuration is established in special quasi-random structures via ATAT package \cite{VANDEWALLE201313}, then fully relaxed and optimized for every lattice constant. We use volumetric-energy method in determination of lattice constant. Details of calculations in lattice constant are referred to the previous works \cite{YOU2024103919, YOU2025104187}. After determining the lattice constants, we apply a slip at a different atomic plane, and calculate the SFEs. The SFE is calculated as below \cite{Vítek01101968},
\begin{align} \label{eqn:my_equation}
    \gamma=\frac{E-E_{0}}{A}
\end{align}
where $E_{0}$ and $E$ are energies of initial and deformed states, and $A$ is an area of slip-plane. The DFT calculations are implemented within the projected-augmented-waves (PAW) \cite{PhysRevB.59.1758} approach via Perdew-Burke-Ernzerhof (PBE) \cite{PhysRevLett.77.3865} exchange-correlation potentials. The PAW-PBE pseudopotentials for Ni, Co, Cr, Fe, Mn, and V are employed with the following valence electron configurations Ni ($3p^6$ $3d^9$ $4s^1$), Co ($3d^8$ $4s^1$), Cr ($3p^6$ $3d^5$ $4s^1$), Fe ($3p^6$ $3d^7$ $4s^1$), Mn ($3p^6$ $3d^6$ $4s^1$), and V ($3s^2$ $3p^6$ $3d^4$ $4s^1$), respectively. The oriented FCC superlattice with 270 atoms is employed with Monkhorst-pack k-mesh 2 × 2 × 2 for electronic minimization and ionic relaxation in allowing full distortions and lattice volumetric optimization, and 2 × 2 × 1 for determining SFEs to hinder redundant SFs outside the periodic boundary. A 450 eV plane-wave energy cut-off is used for all the calculations. The tolerance criteria for energy and force are allowed within 1 meV and 5 meV/Å, respectively. For SFE of each alloy at finite temperatures, we employ ab-initio molecular dynamics (AIMD) \cite{YOU2024103919, YOU2025104187}. This approach does not rely on 1) the empirical interatomic potential that is commonly used in the classical MD, and 2) the extrapolation of free energy from 0 K (such as quasi-harmonic approximation), so that we can more accurately estimate the SFEs at finite temperatures. For the electronic-minimization in AIMD, we use the same PAW-PBE pseudopotentials from the DFT calculation. A Nosé-Hoover thermostat is chosen for drawing an equilibrium state from a canonical ensemble of sampled states \cite{PhysRevA.31.1695, 10.1063/1.447334}. This introduces fictitious friction terms to permit temperature control into the Hamiltonian such that,
\begin{align} \label{eqn:my_equation}
H'=\sum\limits_{i=1}^N \frac{m_{i}}{2}s^2(\dot{\overrightarrow{r_{i}}})^2 -U_{p}(\overrightarrow{r})+\frac{Q}{2}\dot{s}^2+3NkT\ln (s)
\end{align}
where $N$ is the total number of atoms, $m$ the mass of atom in $i$ site, $s$ the fictitious parameter introduced, $\overrightarrow{r}$ the position of atoms, so that the first two terms are kinetic and potential energy ($U_{p}$) of the given system. $Q$ is the effective mass of $s$, $k$ is the Boltzmann constant, and $T$ is the temperature. We set $Q=1$, and the simulation is controlled in 2 fs time-step (total 600 $\sim$ 700 steps, i.e., 1.2 $\sim$ 1.4 ps), 450 eV plane-wave energy cut-off, and single k-point $\Gamma$ mesh-grid. 
We use the optimized structures at 0 K as initial points. Among the multiple slip-cuts, we only select the one slip-cut that gives the maximum SFEs in each slip system, since it dictates the macroscopic deformation \cite{YOU2024103919, YOU2025104187, YOU2025113175}. On the slip-cut of the maximum SFE, we set the temperatures of 77 K and 300 K for each alloy. The equilibrated temperature along the time frame approaches the target temperature on average for all the calculations. The proper equilibrating step is determined by the minimum error of mean, $\epsilon$. The error of mean at each cutoff step ($i$) is defined by $\epsilon_{i}=\sqrt{\frac{V_{i}}{N_{t,i}\kappa}}$, where $V_{i}$ is variance, $N_{t,i}$ the number of data at each cutoff, and $\kappa$ is correlation time based on autocorrelation such that,
\begin{align} \label{eqn:my_equation}
\kappa=1+2\sum\limits_{j=1}^{N_{t}} \frac{1}{V(N_{t}-j)} \sum\limits_{k=1}^{N_{t}-j} (X_{k}-\overline{X})(X_{k+j}-\overline{X})
\end{align}
where $V$, $\overline{X}$, and $N_{t}$ are variance, average, and total number of data for original $X$ without cutoff. Total energies of undeformed or deformed structures have the converged value on average by excluding the data in non-equilibrating steps.

%\section*{Data Availability}
%The data that support the findings of the present study are available in the data repository Dryad \cite{????/}.

\bibliography{References}

\section*{Acknowledgments}

J.C.S., D.A., M.C. and S.S. acknowledge the National Science Foundation (NSF) DMR CAREER Award \#2338346 for research funding. H.S. and D.Y acknowledge the National Science Foundation (NSF) CMMI Award \#21-25821 for research funding. This work was carried out in the Materials Research Laboratory Central Research Facilities, University of Illinois. M.H. acknowledges financial support from the Czech Science Foundation under the Junior Star project with contract No. 24-11058M. The use of CEITEC Nano Research Infrastructure supported by CzechNanoLab project LM2023051 funded by MEYS CR is gratefully acknowledged.

\section*{Author information}

\section*{Ethics declarations}

The authors declare no competing financial or non-financial interests.

\noindent\textbf{CRediT authorship contribution statement} \\

\textbf{D.A}: Conceptualization, Data curation, Formal analysis, Investigation, Methodology, Resources, Writing – original draft, Writing – review \& editing. \textbf{M.H.}: Investigation, Methodology, Verification, Writing – review \& editing. \textbf{D.Y}: Formal analysis, Resources, Writing – original draft, Writing – review \& editing. \textbf{M.C.}: Verification, Resources, Writing – review \& editing. \textbf{S.S.}: Investigation, Verification, Writing – review \& editing. \textbf{M.R}: Resources. \textbf{A.S.T}: Resources. \textbf{H.S.}: Resources, Verification, Writing – review \& editing. \textbf{G.L}: Resources, Writing – original draft, Writing – review \& editing. \textbf{J.C.S.}: Conceptualization, Funding acquisition, Methodology, Project administration, Resources, Supervision, Writing – original draft, Writing – review \& editing. \\

\newpage
\hypertarget{supplement}{}
\section*{Supplementary Materials}

\thispagestyle{empty}

% \noindent Please note: Abbreviations should be introduced at the first mention in the main text – no abbreviations lists. Suggested structure of main text (not enforced) is provided below.

\renewcommand{\thefigure}{S\arabic{figure}}

\begin{figure}[H]
  \centering
 \includegraphics[width=1\textwidth]{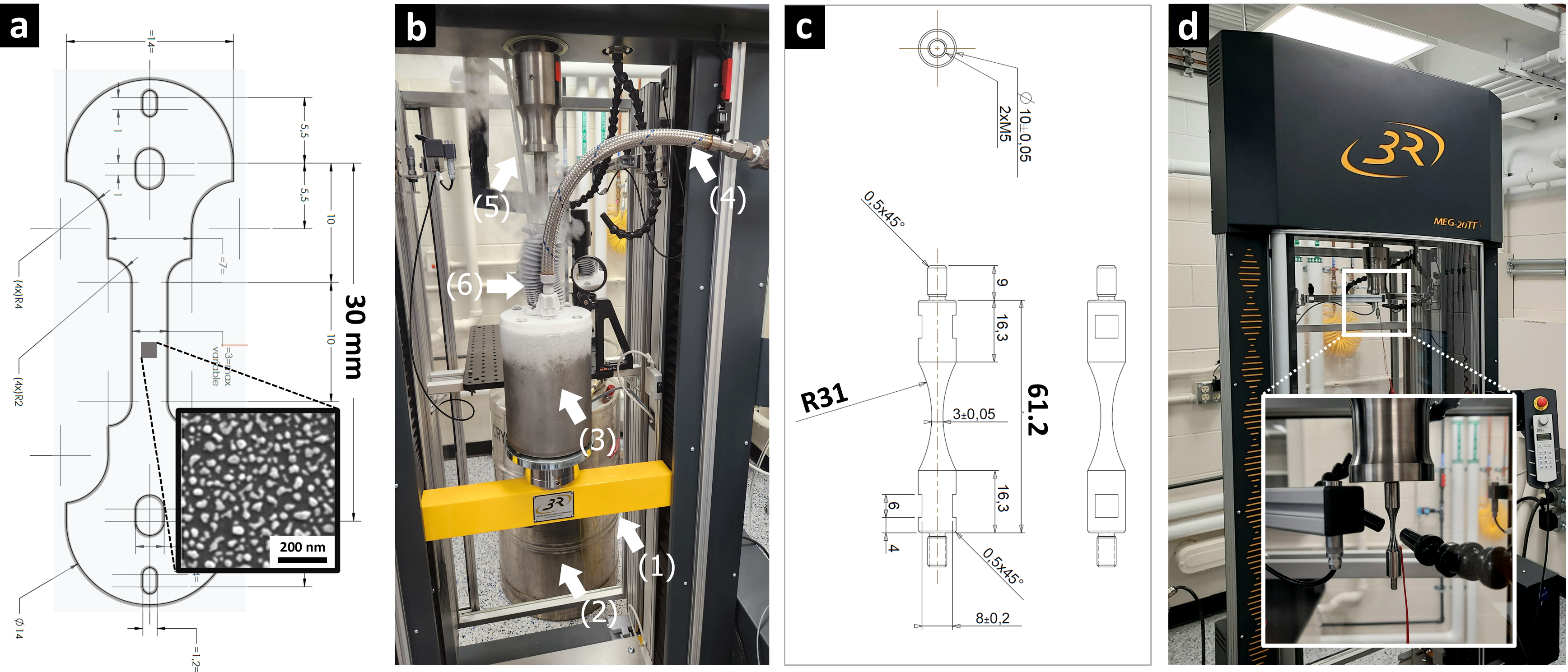}
\caption{\textbf{(a)} Specimen geometry used for mechanical testing and HR-DIC measurements. The corresponding high resolution speckle pattern used for HR-DIC measurements is shown in the inset. \textbf{(b)} Experimental setup for mechanical testing at cryogenic temperatures. (1) Moving traverse for mechanical loading equipped with a 20kN cell force; (2) liquid helium dewar; (3) vacuum double-walled chamber with thermal insulation; (4) recycling line for helium gas; (5) liquid helium transfer line; (6) impervious flexible to seal chamber. \textbf{(c)} Specimen geometry used for very high cycle fatigue (VHCF) testing. Dimensions are provided in millimeters. \textbf{(d)} Experimental setup for VHCF testing. The figure inset shows an enlarged view of the fatigue specimen with a strain gauge attached to record the applied strain during cyclic loading.}
\label{Setup}
\end{figure}

Fig. \ref{Setup}(a) shows the specimen geometry of the tensile specimens used for monotonic mechanical testing. All dimensions are provided in millimeters and the specimens were approximately 1 mm in thickness. The inset in \textbf{(a)} depicts a magnified secondary electron image of the high resolution speckle pattern used for HR-DIC measurements. The speckle pattern is generated by the reconfiguration of a 10 nm silver layer and remains stable throughout the range of considered temperatures (ambient and cryogenic). Fig. \ref{Setup}(b) shows the experimental setup for cryogenic tensile testing. Mechanical tests can be performed either in a liquid nitrogen or liquid helium environment. Specimens are immersed in liquid nitrogen or helium and the liquid level is maintained by controlling the flow using nitrogen or helium gas, respectively. Fig. \ref{Setup}(c) illustrates the specimen geometry of the fatigue specimens used for very high cycle fatigue (VHCF) testing. Dimensions are provided in millimeters and the specimens are designed to resonate at a frequency of 20 kHz during cyclic loading. Fig. \ref{Setup}(d) presents the experimental setup for VHCF testing. Specimens are mounted onto a 3R$\texttrademark$ VHCF MEG20TT ultrasonic fatigue testing frame which operates at a testing frequency of 20 kHz. The inset in \textbf{(d)} displays an enlarged view of the fatigue specimen. A compressed air flow is directed onto the specimen to eliminate self-heating. A strain gauge is attached to the gauge section of the specimen to record the applied strain during cyclic loading and estimate the corresponding stress level.

\newpage

\begin{figure}[H]
  \centering
 \includegraphics[width=1\textwidth]{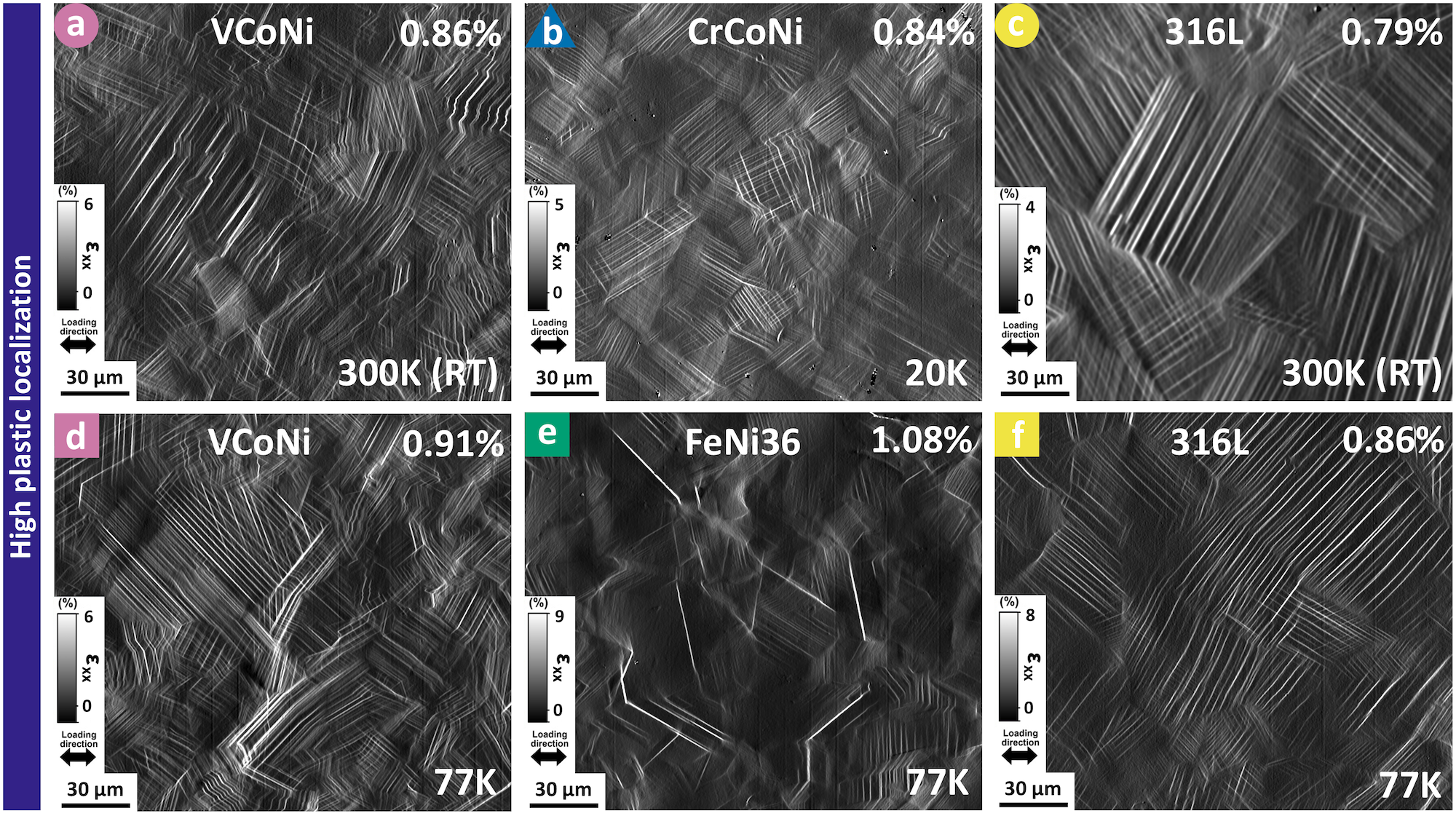}
\caption{HR-DIC $\epsilon_{xx}$ longitudinal strain maps for \textbf{(a)} VCoNi deformed at room temperature, \textbf{(b)} CrCoNi deformed at 20K, \textbf{(c)} stainless steel 316L deformed at room temperature, \textbf{(d)} VCoNi deformed at 77K, \textbf{(e)} FeNi36 deformed at 77K and \textbf{(f)} stainless steel 316L deformed at 77K. The macroscopic plastic strains up to which all materials were deformed are noted on the respective HR-DIC maps. All materials were deformed in tension and the tensile direction is horizontal.}
\label{HighLocalization}
\end{figure}

Fig. \ref{HighLocalization} shows reduced regions of the HR-DIC $\epsilon_{xx}$ longitudinal strain maps for a selection of materials and testing conditions investigated in this study. The materials, testing temperature and macroscopic plastic strains up to which they were deformed are included on the individual HR-DIC maps. This selection of maps corresponds to the plastic deformation response referred to as "High" plastic localization. The characteristic feature of this plastic localization response is the occurrence of individual, discrete and intense deformation events developing as a result of plastic deformation. This plastic response, typically shown by most FCC metals and alloys, is attributed to the glide of dislocations on discrete crystallographic planes, resulting in pronounced extrusions on the material surface. A direct consequence of this response is a relatively high plastic localization intensity corresponding to these deformation events. The plastic localization intensity is expressed in nanometers and corresponds to the in-plane displacement induced by each of these deformation events. This metric is obtained from HR-DIC measurements and enables the quantification of the material's plastic response. 

\newpage

\begin{figure}[H]
  \centering
 \includegraphics[width=1\textwidth]{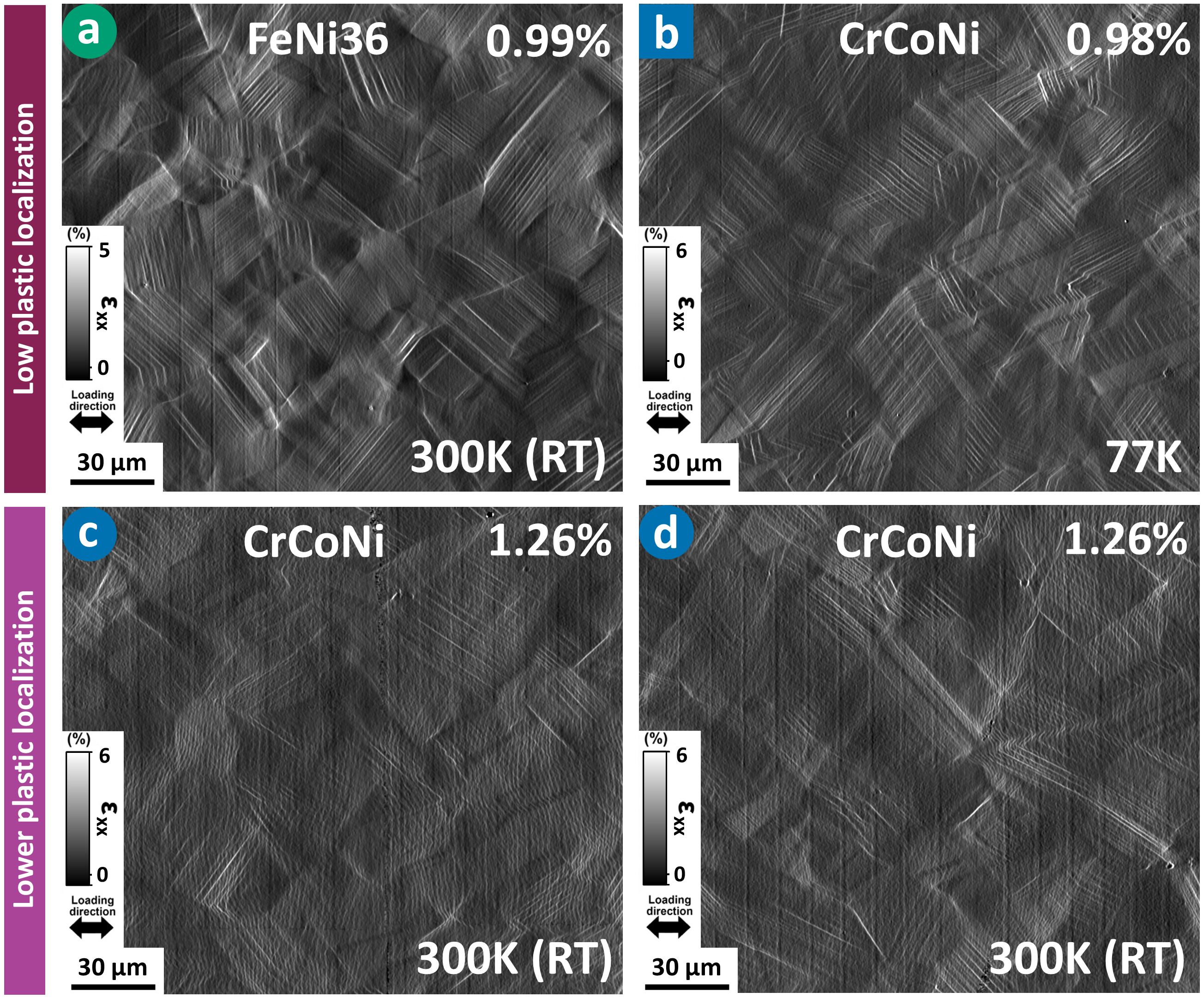}
\caption{HR-DIC $\epsilon_{xx}$ longitudinal strain maps for \textbf{(a)} FeNi36 deformed at room temperature, \textbf{(b)} CrCoNi deformed at 77K and \textbf{(c,d)} CrCoNi deformed at room temperature. The macroscopic plastic strains for the tested materials are noted on the HR-DIC maps. All materials were deformed in tension and the tensile direction is horizontal.} %The displayed maps in \textbf{(a)} and \textbf{(b)} correspond to the plastic localization response termed as "Low", which leads to a reduction in the plastic localization intensity expressed in nanometers, while maps displayed in \textbf{(c)} and \textbf{(d)} correspond to the plastic localization response termed as "Lower", which leads to an even greater reduction in the plastic localization intensity. Both of these responses are characterized by homogeneous, diffused plasticity which is in contrast to the localized, intense plasticity observed for conventional FCC metals as seen in Fig. \ref{HighLocalization}.}
\label{LowLowerLocalization}
\end{figure}

Fig. \ref{LowLowerLocalization} represents reduced regions of the HR-DIC $\epsilon_{xx}$ longitudinal strain maps for a selection of materials and testing conditions investigated in this study. The materials, testing temperature and macroscopic plastic strains up to which they were deformed are included on the individual HR-DIC maps. The maps displayed in \textbf{(a)} and \textbf{(b)} correspond to the plastic deformation response referred to as "Low" plastic localization. Compared to the "High" plastic localization response, the plastic localization intensity or the in-plane displacement induced by deformation events is reduced. Similarly, the maps displayed in \textbf{(c)} and \textbf{(d)} correspond to the plastic deformation response referred to as "Lower" plastic localization. The plastic localization intensity in these cases is even lower than the "Low" localization case. As evidenced from the HR-DIC maps, both the "Low" and "Lower" plastic localization responses are characterized by homogeneous, diffused plasticity, extending over entire crystallographic grains in many cases. This particular effect is even more pronounced in \textbf{(c)} and \textbf{(d)}, leading to the lowest plastic localization. This is in contrast to the "High" plastic localization response, which is characterized by localized, individual and intense deformation events, as observed in Fig. \ref{HighLocalization}.

\newpage

\begin{figure}[H]
  \centering
 \includegraphics[width=1\textwidth]{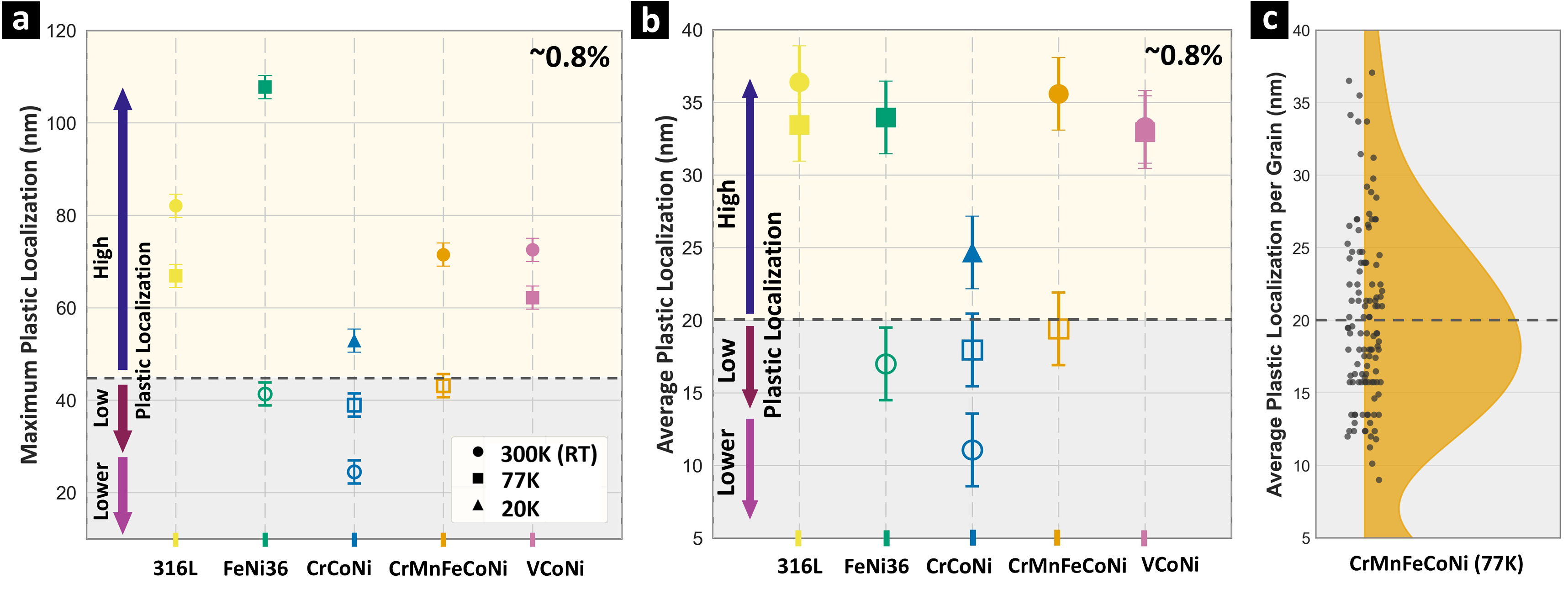}
\caption{\textbf{(a)} 5\% highest maximum plastic localization and \textbf{(b)} average plastic localization expressed in nanometers for the investigated materials in this study deformed up to a macroscopic plastic strain of approximately 0.8\%. \textbf{(c)} Distribution of the average plastic localization per grain expressed in nanometers for CrMnFeCoNi deformed at 77K upto a plastic strain of 0.83\%. The data points in \textbf{(c)} have been spread out along the x axis to enhance visual clarity.}
\label{MaxBlN}
\end{figure}

Fig. \ref{MaxBlN}(a) presents the 5\% highest maximum plastic localization and Fig. \ref{MaxBlN}(b) shows the average plastic localization for the investigated materials in this study. The plastic localization is expressed in nanometers and corresponds to the in-plane displacement induced by deformation events. The materials were deformed up to a macroscopic plastic strain of approximately 0.8\% and tested at room (300K), liquid nitrogen (77K) and near liquid helium temperature (20K). Arbitrary thresholds are defined at 45 nm and 20 nm in \textbf{(a)} and \textbf{(b)} respectively. These thresholds, illustrated as horizontal dashed lines, classify materials based on their plastic deformation response as "High" and "Low", "Lower" plastic localization. Materials developing intense, localized plasticity in the form of discrete deformation events are represented on a yellowish background above the defined threshold, whereas materials developing homogeneous, diffused plasticity are displayed on a gray background below the threshold. Fig. \ref{MaxBlN}(c) represents the distribution of the average plastic localization per grain for the high entropy alloy CrMnFeCoNi deformed at 77K up to a macroscopic plastic strain of 0.83\%. Each data point in the distribution corresponds to the average value of the plastic localization intensity for all detected deformation events within an individual grain. This material when deformed at 77K, lies just below the defined thresholds in \textbf{(a)} and \textbf{(b)} as evidenced by the distribution of points centered about 20 nm, indicating that grains developing both high and low localization are present as a result of plastic deformation. 

\newpage

\begin{figure}[H]
  \centering
 \includegraphics[width=1\textwidth]{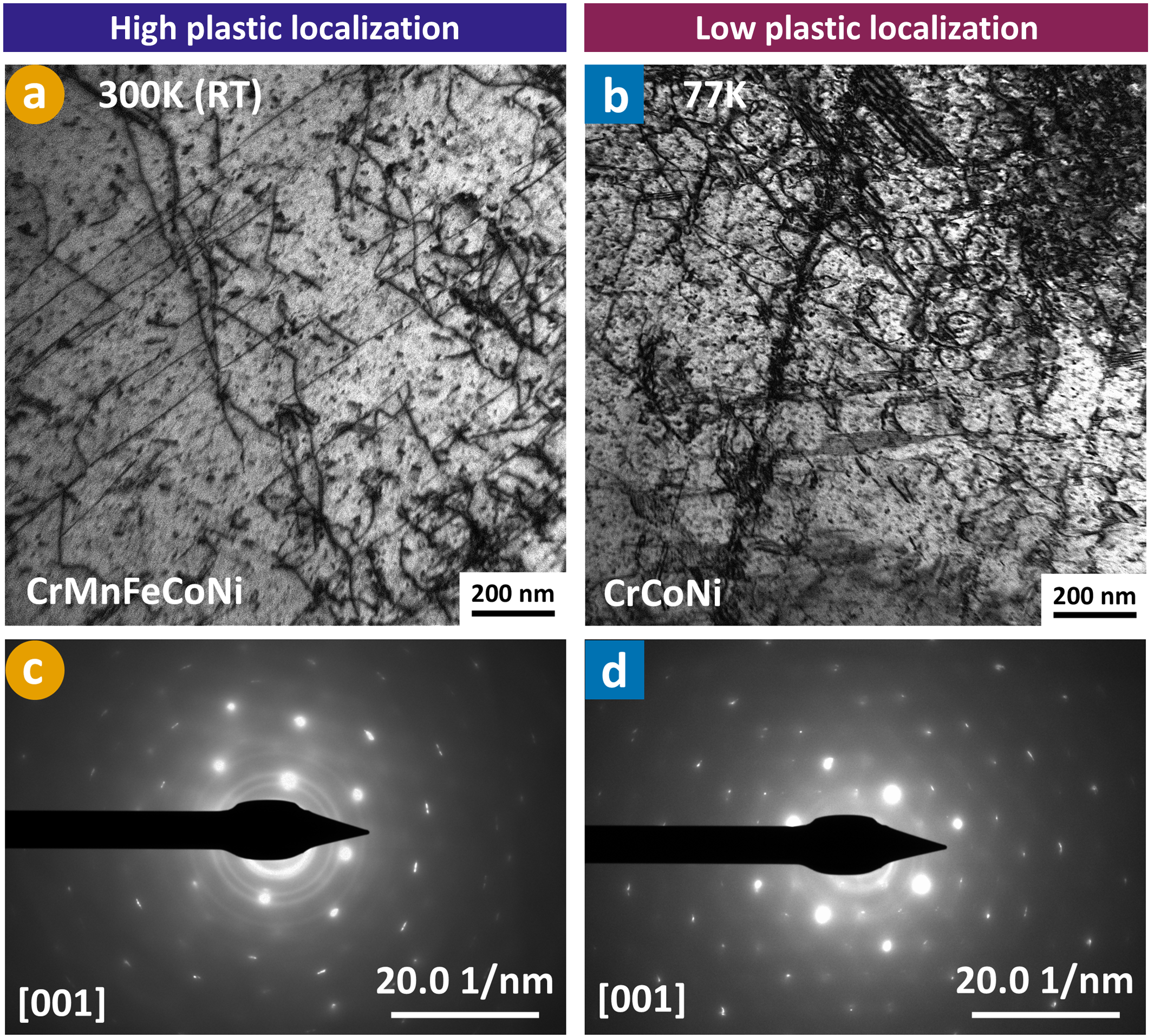}
\caption{Bright-field (BF) TEM images of thin foils extracted from \textbf{(a)} the high entropy alloy CrMnFeCoNi deformed at 300K up to 0.95\% of plastic strain and \textbf{(b)} the medium entropy alloy CrCoNi deformed at 77K up to 0.47\% of plastic strain. The selected area diffraction patterns (SADPs) corresponding to \textbf{(a)} and \textbf{(b)} are shown in \textbf{(c)} and \textbf{(d)} respectively.}
\label{Diffraction}
\end{figure}

Fig. \ref{Diffraction}(a) and (b) show bright-field (BF) transmission electron microscopy (TEM) images of thin foils extracted from the high entropy alloy CrMnFeCoNi deformed at room temperature (300K) and CrCoNi deformed at 77K, respectively. The corresponding selected area diffraction patterns (SADPs) after aligning the foils along the [001] zone axis are shown in \textbf{(c)} and \textbf{(d)}. CrMnFeCoNi deformed at 300K shows the plastic deformation response termed as "High" plastic localization, characterized by full dislocation glide and corresponding to the deformation response in conventional FCC metals, as is evident from the SADP in \textbf{(c)} which illustrates diffraction spots indicating the FCC matrix. CrCoNi deformed at 77K develops the plastic deformation response termed as "Low" plastic localization. In addition to dislocations, extensive planar defects such as stacking faults are evident in the BF TEM image. The corresponding SADP in \textbf{(d)} demonstrates additional diffraction spots apart from the FCC matrix and suggests the presence of other planar defects such as deformation twins forming at the nanometer scale and not directly evident in the BF TEM image, highlighting differences between the different plastic deformation responses.

\newpage

\begin{figure}[H]
  \centering
 \includegraphics[width=1\textwidth]{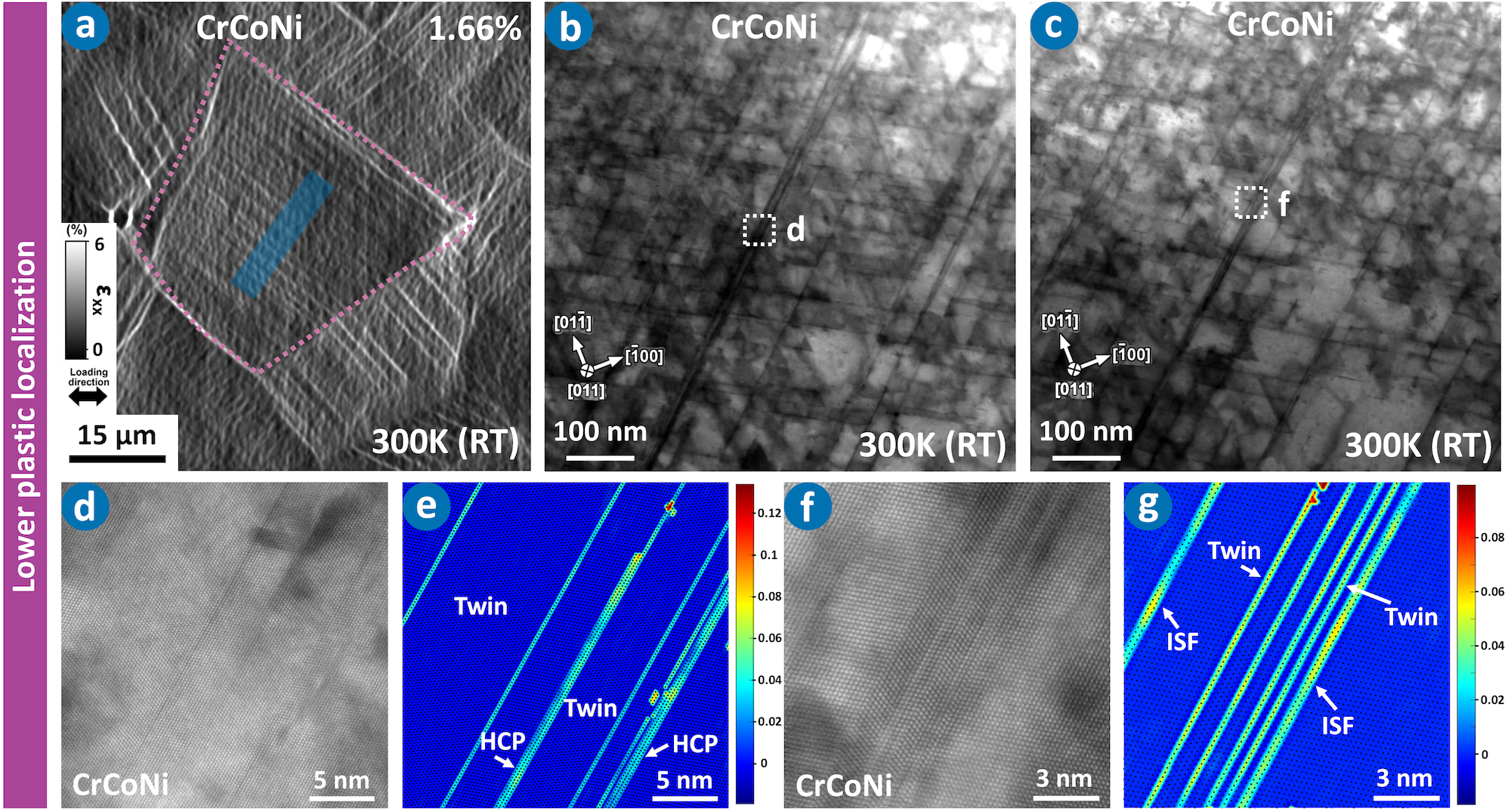}
\caption{\textbf{(a)} HR-DIC $\epsilon_{xx}$ longitudinal strain map for CrCoNi deformed at room temperature. The extracted thin TEM foil from the considered grain of interest is marked by a blue region. The associated bright field STEM DCI micrographs are displayed in \textbf{(b)} and \textbf{(c)}. Atomic resolution HAADF-STEM images corresponding to reduced regions of interest in \textbf{(b,c)} and shown in \textbf{(d)} and \textbf{(f)}, respectively. The associated center of symmetry (COS) analyses in \textbf{(e,g)} correspond to \textbf{(d)} and \textbf{(f)}, respectively. The electron beam was parallel with the [011] crystallographic zone axis in \textbf{(b-g)}.}
\label{TEM2_Supplementary}
\end{figure}

Fig. \ref{TEM2_Supplementary}(a) shows a reduced region of the HR-DIC $\epsilon_{xx}$ longitudinal strain map for CrCoNi deformed at room temperature with a grain of interest. The thin TEM foil extracted from this grain is represented by a blue region. CrCoNi when deformed at room temperature is associated with the plastic localization response referred to as "Lower" plastic localization. This response is characterized by homogeneous and diffused plasticity, leading to drastically low values of the plastic localization intensity in nanometers. The associated bright-field STEM DCI micrographs of the foil are displayed in \textbf{(b)} and \textbf{(c)}. These images reveal deformation events with extended thicknesses. Atomic resolution HAADF-STEM images corresponding to the reduced regions marked by white dashed boxes in \textbf{(b)} and \textbf{(c)} are shown in \textbf{(d)} and \textbf{(f)}, respectively. Center of symmetry (COS) analyses for defect identification on the atomic resolution images reveal that these deformation events with extended thicknesses consist of a high density of planar defects, dominated by deformation nanotwins. This emphasizes the role of planar defects and deformation nanotwins in the observed homogeneous plasticity response.

\newpage

\begin{figure}[H]
  \centering
 \includegraphics[width=1\textwidth]{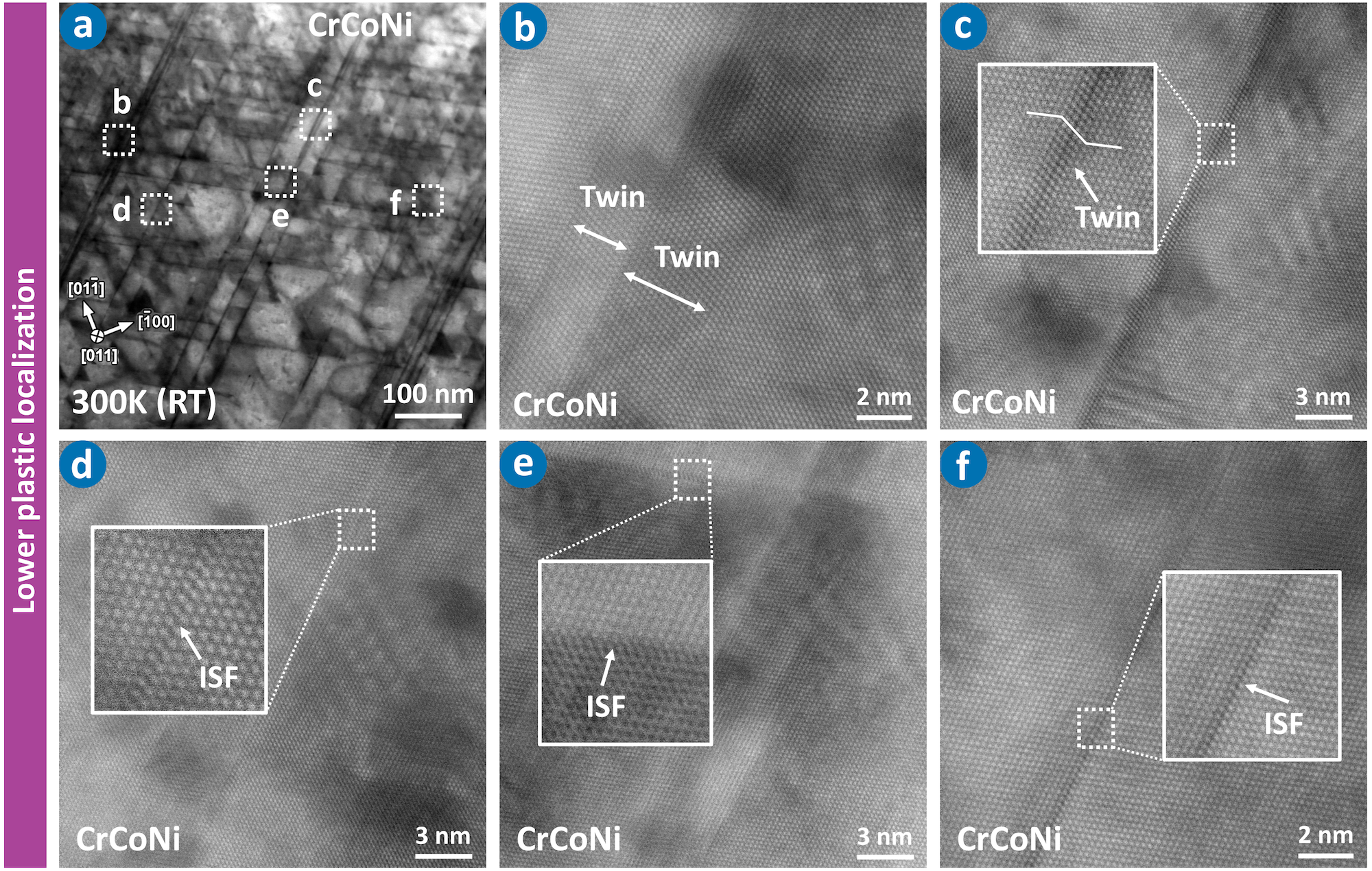}
\caption{\textbf{(a)} Bright field STEM DCI micrograph revealing a deformation microstructure in a foil extracted from a grain developing homogeneous plasticity for CrCoNi deformed at room temperature. Atomic resolution HAADF-STEM images for deformation events displaying an extended thickness are shown in \textbf{(b)} and \textbf{(c)}, whereas atomic resolution HAADF-STEM images corresponding to regions outside or between these extended events are presented in \textbf{(d)}, \textbf{(e)} and \textbf{(f)}, respectively. The atomic resolution images correspond to the locations marked by white dashed boxes marked in \textbf{(a)}. White dashed boxes in \textbf{(c-f)} are enlarged in the figure insets to highlight specific planar defects. The electron beam was parallel with the [011] crystallographic zone axis.}
\label{TEM3_Supplementary}
\end{figure}

Fig. \ref{TEM3_Supplementary}(a) shows a bright field STEM DCI micrograph of a thin foil extracted from a grain showing homogeneous, diffused plasticity for the medium entropy CrCoNi alloy deformed at room temperature. This alloy is associated with the plastic localization response referred to as "Lower" plastic localization at 300K, which leads to a drastic reduction in the plastic localization intensity. The displayed BF-STEM DCI micrograph reveals the presence of deformation events that exhibit an extended thickness. Atomic resolution HAADF-STEM images for two such events corresponding to white dashed boxes in \textbf{(a)} are displayed in \textbf{(b)} and \textbf{(c)}. It is observed that these deformation events are dominated by the presence of deformation nanotwins, as is evident from the HAADF-STEM images. The corresponding atomic resolution HAADF-STEM images for regions within the TEM foil, outside of, or in between these extended deformation events are given in \textbf{(d)}, \textbf{(e)} and \textbf{(f)}. These images also correspond to the white dashed boxes marked in \textbf{(a)}. For these specific locations, it is observed that other planar defects, mainly intrinsic stacking faults are evident in the HAADF-STEM images. This reveals the contrasting types of planar defects distributed outside deformation events with extended thicknesses, further highlighting the role of deformation nanotwins contributing to greater homogenization of plasticity.

\newpage

\begin{figure}[H]
  \centering
 \includegraphics[width=1\textwidth]{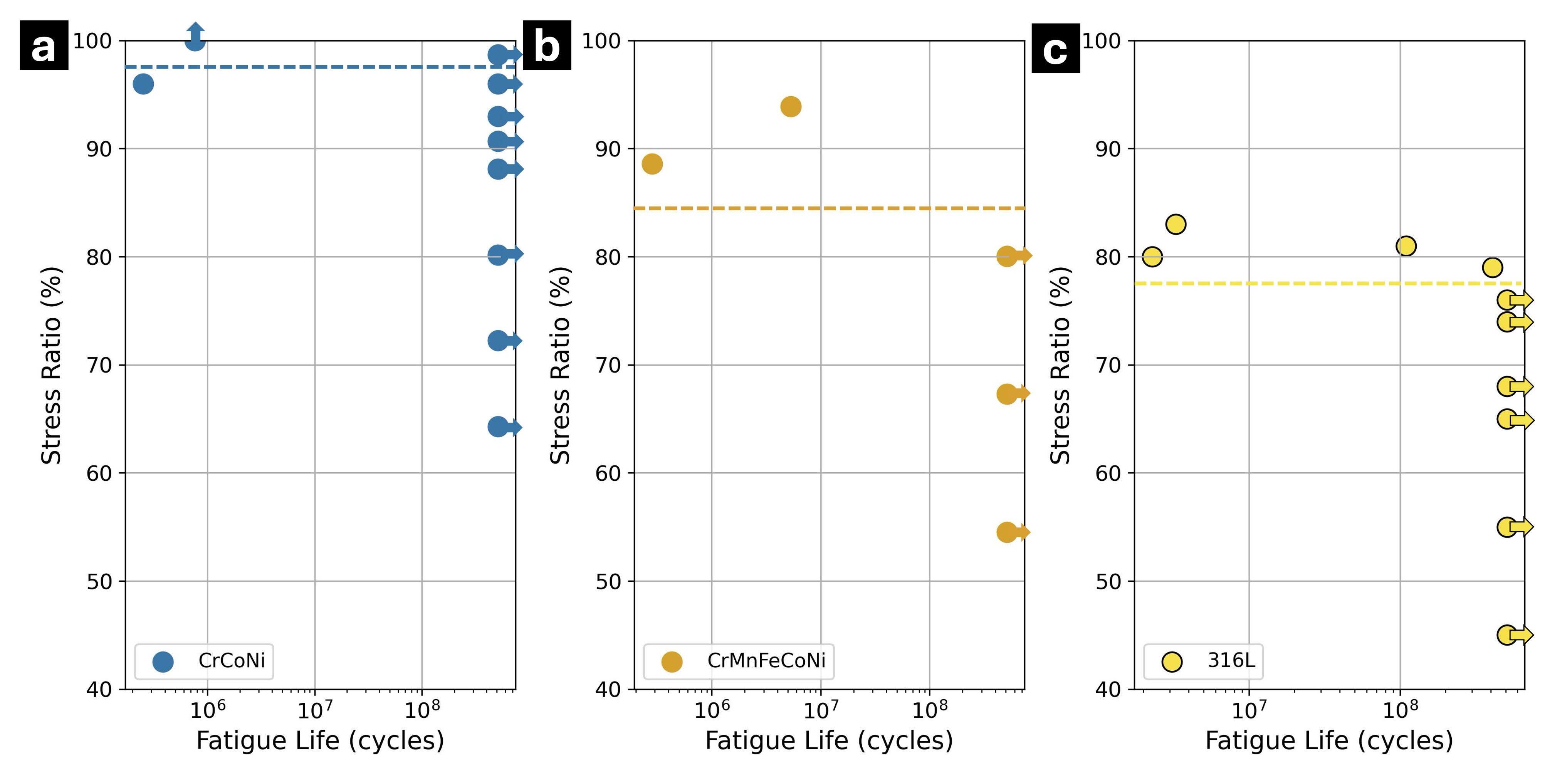}
\caption{Fatigue life at room temperature (300K) for \textbf{(a)} CrCoNi, \textbf{(b)} CrMnFeCoNi and \textbf{(c)} stainless steel 316L alloys investigated in this study as a function of the maximum applied stress in the very high cycle fatigue (VHCF) regime. The stress ratio corresponds to the ratio of the maximum applied stress and the yield strength of the material, expressed in percentage. Horizontal arrows represent specimens that did not fail at a given stress level. Vertical arrows indicate cases where the specimens have a fatigue life that exceeds the yield strength. The fatigue ratio is represented using horizontal dashed lines.}
\label{Fatigue_Supplementary}
\end{figure}

Fig. \ref{Fatigue_Supplementary} shows the fatigue life at room temperature for the investigated alloys in this study, namely \textbf{(a)} CrCoNi, \textbf{(b)} CrMnFeCoNi and \textbf{(c)} stainless steel 316L. The fatigue testing was conducted in the very high cycle fatigue regime (VHCF). The fatigue life is represented using the fatigue ratio, which corresponds to the ratio of the maximum applied stress and the yield strength of the material. Specimens were subjected to successively increasing stress levels until failure. The specimens that did not fail at a given applied stress are shown using horizontal arrows. Vertical arrows indicate cases where the specimens had a fatigue ratio greater than 100\%, corresponding to a fatigue life exceeding the yield strength. The obtained fatigue ratio is represented using horizontal dashed lines. Based on previously reported correlations between the fatigue strength and intensity of plastic localization, intense localization of plasticity leads to a lower fatigue life and vice versa. CrMnFeCoNi and stainless steel 316L tested at room temperature that correspond to the plastic localization response termed as "High" are compared with CrCoNi deformed at room temperature, which exhibits the plastic localization response termed as "Lower", characterized by homogeneous plasticity and significantly lower intensity of localization. The observed fatigue results are in agreement with the correlation between localization intensity and fatigue strength, with CrCoNi exhibiting a considerably higher fatigue ratio as compared to CrMnFeCoNi and 316L. These results highlight the role of the observed dynamic plastic deformation delocalization mechanism in improving critical long term mechanical properties such as the fatigue strength.

\end{document}